%% file: iclr2027_conference.tex
\documentclass{article} 
\usepackage{iclr2027_conference,times}

\input{math_commands.tex}

\usepackage[utf8]{inputenc} 
\usepackage[T1]{fontenc}    
\usepackage{hyperref}
\usepackage{url}
\usepackage{booktabs}       
\usepackage{graphicx}       
\usepackage{adjustbox}      
\usepackage{nicefrac}       
\usepackage{microtype}      
\usepackage{xcolor}         
\usepackage{algorithm}
\usepackage{algpseudocode}
\usepackage{subcaption}     

\title{\textit{ConventionPlay}: Capability-Limited Training for Robust Ad-Hoc Collaboration}

\author{Abhishek Sriraman, Eleni Vasilaki \& Robert Loftin \\
Department of Computer Science\\
Sheffield, UK \\
Pittsburgh, PA 15213, USA \\
\texttt{\{asriraman1,e.vasilaki,r.loftin\}@sheffield.ac.uk} \\
}

\iclrfinalcopy 
\begin{document}
\include{iclr_data/convplay_tables}
\include{iclr_data/rebalance_tables}
\include{iclr_data/hyperparams_tables}

\maketitle

\begin{abstract}
Ad-hoc collaboration often requires agents to identify and adhere to some shared convention within a cooperative task.  Existing work on reinforcement learning (RL) for ad-hoc collaboration focuses on training agents that adapt to the conventions established by their partners. These methods fail to consider the possibility that while some partners might follow only a single fixed convention, others may themselves be capable of adapting to multiple conventions.  Here we present \textit{ConventionPlay}, an RL-based approach that teaches agents to discover their partner's optimal convention by training against a learned population of partners that exhibit different degrees of adaptability across conventions.  Some of these partners follow a single, fixed convention, while others are able to adapt to a subset of the possible conventions for the task in question.  The existence of partners that support a limited subset of conventions forces agents trained against this population to actively probe their partner's capabilities, and steer their partner towards the most effective joint strategy that they are capable of following.  Our experimental results demonstrate that agents trained via \textit{ConventionPlay} achieve superior performance to existing ad-hoc collaboration methods against test populations of partners that are compatible with multiple conventions.
\end{abstract}

\section{Introduction}

Cooperative multi-agent tasks often admit a variety of different solutions, with success depending on the agents' ability to coordinate their strategies on a single solution.  In such tasks, joint strategies can exhibit different levels of compatibility, with performance deteriorating when individual strategies are misaligned.  We describe collections of mutually compatible strategies as ``conventions''.  Within the broader challenge of designing agents to collaborate with unseen teammates \citep{mirsky2022surveyadhocteamwork}, the ability to recognize and align with established conventions is crucial for effective ad-hoc collaboration \citep{Stone_Kaminka_Kraus_Rosenschein_2010,DBLP:conf/iclr/ShihSKES21,10.1145/3306618.3314268}.  In this work, we propose an approach to training collaborative agents for a novel setting -- interacting with partners that have varying levels of proficiency over known conventions.

As a motivating example, consider an agent joining a cooperative setting such as a mechanic shop, a football pitch, or a professional kitchen.  In each of these domains team members may possess a diverse and partially overlapping set of skills---e.g. in a kitchen, most chefs know how to both chop and sauté vegetables, but only some may be well versed in using a specialized tool.  The skills of the individual agents determine the set of joint strategies available to the team, one of which they must actually commit to.  A new chef joining the team then has two options.  They could infer the status quo, and adapt to an established convention. Alternatively, they could act on an opportunity to produce a better outcome for the team by changing the joint strategy---for instance, reassigning tasks among the cooks---in the hope that the rest of the team would be able to adapt accordingly. 

We consider how prior methods for ad-hoc collaboration would approach this scenario. A common approach \citep{strouseFCPCollaboratingHumans2021, zhaoMEPMaximumEntropy2022, carroll2019utility} trains collaborative agents through a two-step process: First, a diverse ``population'' of agents is trained for the target task---e.g. teams of cooks where each team assumes a certain task allocation. Then, a ``generalist'' agent is trained to coordinate with any member of that population.  Each member of the initial population effectively acts as a ``leader,'' committed to a fixed convention and expecting its partner to adapt to it.  The generalist agent trained against these partners then becomes a ``follower'' that adapts to its partner's convention.  A shortcoming here is that the generalist defers unconditionally to its partner's convention, however sub-optimal, since it has never encountered a partner who could themselves adapt to a different convention.  This does not reflect realistic populations, where partners vary in their capacity for adaptation---some follow only one convention, others several. Because existing methods do not model this variation, the resulting agents remain purely reactive, with no mechanism for recognizing when a partner could instead be steered toward a higher-value convention.

In this work, we introduce \textit{ConventionPlay}, a reinforcement learning-based approach to training adaptive agents that are capable of coordinating with partners that may themselves be adaptive.  The novelty of this algorithm lies in discovering how a population partitions into conventions, and using this partition to construct a structured training population of \textit{generalist followers}---agents designed to adapt to specific, restricted subsets of conventions.  Exposing our agent to these adaptive partners forces it to move beyond purely reactive behavior; it must actively probe its partner to determine whether to steer the team toward a higher-value convention, or defer to the partner's established one. Our experimental results demonstrate that \textit{ConventionPlay} agents outperform state-of-the-art methods \citep{strouseFCPCollaboratingHumans2021,cuiSyKLRBRKlevelReasoning2021,sarkarComediDiverseConventions2023} in identifying the most effective convention its partner can follow in a number of complex ad-hoc collaboration scenarios.

\section{Related Work}
\label{sec:related_work}

\paragraph{Population Diversity} The most common method for achieving robust ad-hoc collaboration is by anticipating the range of possible solutions to the task.  A number of methods focus on computing a diverse set of solutions and training a generalist agent against this set of strategies.  Fictitious Co-Play (FCP) \citep{strouseFCPCollaboratingHumans2021} provides behavioral diversity using intermediate checkpoints from a self-play training process, simulating partners of varying skill levels.  \citet{carroll2019utility} extract diversity from human interaction data in an offline dataset.  TrajeDI \citep{lupuTrajectoryDiversityZeroShot2021} incentivizes policies to generate distinct trajectories, while MEP \citep{zhaoMEPMaximumEntropy2022} uses an entropy bonus to encourage the population to cover the entire state-action space. LIPO \citep{charakorn2023generating}, CoMeDi \citep{sarkarComediDiverseConventions2023} and ADVERSITY \citep{cuiADVERSARIALDIVERSITYHANABI2023} aim to learn policies that are incompatible with each other, yet individually optimal in self-play. Our work focuses not on discovering the task's solutions, but on leveraging these solutions to train a generalist agent that coordinates with partners of varying proficiency. To that end, our proposed method is compatible with any of the diversity-generation techniques discussed above.  

We contrast our approach against two additional methods that structure diversity hierarchically. TBS \citep{10.65109/UMCV8852} analyzes population diversity with a return-based similarity measure to train an ensemble of best-response agents and select among them online via a theory-of-mind classifier. SyKLRBR \citep{cuiSyKLRBRKlevelReasoning2021} implements a cognitive hierarchy, training agents at each level as best-responses to a uniform distribution over all lower levels. Neither method, however, considers the range of adaptive behaviors a real-world partner may exhibit.

\paragraph{Convention-Free Strategies} Unlike methods that generate diverse solutions, another line of work focuses on zero-shot coordination (ZSC), aiming to avoid conventions that an independently-trained partner is unlikely to discover on its own.  Other-Play \citep{huOtherPlay2020} addresses the equilibrium selection problem by computing ``focal'' policies that are invariant to task symmetries.  We argue this is unrealistic for human-AI coordination: humans do not always converge on such focal policies even when they exist, instead settling on conventions shaped by personal context and prior interactions.  Off-Belief Learning \citep{pmlr-v139-hu21c} suppresses \textit{arbitrary} conventions---e.g. secret handshakes that emerge to convey a partner's private state information in partially observed settings.  Our emphasis is not on partial observability of the environment state, but on the ad-hoc agent's uncertainty about which equilibrium its partner prefers.  Instead of avoiding conventions, our approach embraces them, aiming to construct agents capable of coordinating with any convention present in a population.

\paragraph{Partner Shaping} We distinguish our method from explicit partner-shaping approaches. \citet{pmlr-v97-jaques19a} employ intrinsic rewards to explicitly encourage policies that influence the actions of other agents, rewarding influence itself as an objective.  With our method, influence arises naturally from a best-response objective against a population of partners with latent capabilities, requiring the agent to explore which capabilities a given partner supports.  LOLA \citep{foersterLOLALearningOpponentLearning2018} and M-FOS \citep{luModelFreeOpponentShaping2022} similarly learn strategies for shaping a partner's adaptive behavior.  LOLA however requires a predefined model of the partner's learning process.  While M-FOS does not require such a model, it must repeatedly interact with a specific partner to learn one, with the ability to ``reset'' the partner's learning process multiple times.  Unlike these methods, \textit{ConventionPlay} makes no assumptions about the specific learning algorithm the partner uses to adapt its strategy to different conventions, and learns a strategy that shapes a partner's behavior within a single episode.  Instead, \textit{ConventionPlay} uses the structure of the task itself to identify the behaviors a partner might be capable of implementing, allowing it to quickly identify the best-response to a new partner within a single episode by probing this partner with a limited set of strategies.

\section{Preliminaries}
\label{sec:preliminaries}

In this section we introduce the formalisms and notation needed to describe \textit{ConventionPlay}.  We define Dec-POMDPs, describe the process of training a best-response to a population of agents, and formalize the idea of a system of conventions.

\subsection{Dec-POMDPs}
\label{sec:decpomdp}

We model each coordination task as a Decentralized Partially Observable Markov Decision Process (Dec-POMDP) \citep{oliehoekConciseIntroductionDecentralized2016}, defined by the tuple $\gM = \langle n, \gS, \gA, P, R, \Omega, O \rangle$, where $n$ is the number of agents. $\gS$ is the set of global states, $\gA = \times^{n}_{i=1} \gA_i$ is the joint action space, and $P(s_{t+1}| s_{t}, a_{t})$ defines the transition probabilities to a new state $s_{t+1}$ given the current state $s_t$ and joint action $a_t \in \gA$. The task is fully cooperative, with all agents receiving the same scalar reward $R(s_t, a_t)$. The joint observation $o_t \in \Omega = \times^{n}_{i=1} \Omega_i$ is drawn from the distribution $O(o_t|s_t, a_{t-1})$, with each agent $i$ only observing $o_{i,t}$. An individual policy $\pi_i: \gH_{i} \to \Delta(\gA_i)$ maps the agent $i$'s local action-observation history, $h_{i,t} = (o_{i,0}, a_{i,0}, o_{i,1}, \dots, a_{i,t-1}, o_{i,t}) \in \gH_{i}$, to a probability distribution over its action space.  A \emph{joint policy} $\vec{\pi} = (\pi_1, \ldots, \pi_n)$ specifies the behavior of all agents. Here the goal for the agents is to maximize their expected \textit{episodic} return:
\begin{equation}
J(\vec{\pi}) = \E_{s_0, a_t \sim \vec{\pi}, s_{t+1} \sim P} \left[ \sum_{t=0}^{T} R(s_t, a_t) \right]
\label{eq:objective}
\end{equation}
where $T$ is the episode length.  We develop \textit{ConventionPlay} in symmetric, two-player Dec-POMDPs, with all of our experimental environments being symmetric.  Symmetry means that $\gA_1 = \gA_2$ and $\Omega_1 = \Omega_2$, such that any individual policy can be executed by either agent, and that $J(\pi, \pi') = J(\pi', \pi)$ for all $\pi, \pi' \in \Pi$.  The initial training populations for \textit{ConventionPlay} will consist of policies $\pi \in \Pi$ trained in \emph{self-play} (SP) to maximize the return $J_{SP}(\pi) = J(\pi, \pi)$ receives when paired with itself.  We note that it is straightforward to transform any asymmetric two-agent Dec-POMDP into a symmetric one, by randomly assigning each agent to one of two possible roles in the environment, and allowing each individual policy to condition on the role to which it has been assigned.  Furthermore, while we do not consider settings with more than two agents here, we see no fundamental barriers to extending \textit{ConventionPlay} to tasks with larger numbers of agents.

\subsection{Best-Response Training}
\label{sec:best_resposne_training}

An established paradigm for training agents capable of ad-hoc collaboration is to first train a finite \textit{population} of partner policies $\gP = \{\pi_1, \dots, \pi_m\} \subset \Pi$, and then train a \textit{best-response} policy against this population~\citep{carroll2019utility,strouseFCPCollaboratingHumans2021}.  A best-response to $\gP$ is an individual policy that maximizes expected joint return over partners sampled uniformly from this population:
\begin{equation}
\pi^* \in \text{BR}(\gP) = \arg\max_{\pi \in \Pi} \E_{\pi' \sim \gP} [J(\pi, \pi')]
\label{eq:br}
\end{equation}
This objective ensures that $\pi$ will perform well across a range of different partners.  Here we train all best-response policies using PPO~\citep{schulman2017proximal} with recurrent (LSTM) policy networks that allow the agent to infer its partner's strategy from the interaction history.  Note that such a best-response policy may achieve lower reward against a specific partner than a policy specialized for that single partner, reflecting the cost of adapting online to that partner.

\subsection{Conventions}
\label{sec:conventions}

Intuitively, a ``convention'' is a feature of the joint behavior of two agents, the knowledge of which is sufficient for these agents to coordinate.  For example, the convention that all cars should drive on the right allows vehicles to avoid collisions without knowing where each one is going.  Here we formalize a \textit{system of conventions} as a collection of subsets $\gC = \{C_1, \dots, C_q\}$ of the policy space $\Pi$ (with $\Pi = \bigcup^{m}_{i=1} C_i$).  Each convention is identified with a set $C_i \in \gC$ of strategies that \emph{implement} that convention.  Note that the sets $C_i$ need not be disjoint, as there may be adaptive policies that can implement multiple conventions depending on which partner they are paired with.  A useful system should allow us to determine whether two agents are compatible based solely on the conventions that they implement.  For example, in a task requiring natural language communication between agents, each convention in $\gC$ could correspond to a specific language (e.g., English or French).  Any two policies $\pi$ and $\pi'$ belonging to a single convention $C_i \in \gC$ will be able to successfully cooperate because they speak a common language.  In this work, we define compatibility in terms of return alone, with compatible policies (those that implement compatible conventions) achieving high joint return when paired together, while pairs of incompatible policies yield low returns.  Crucially, all conventions may not be equally good, with some yielding substantially higher joint return than others.  \textit{ConventionPlay} leverages the \textit{cross-play matrix} of its initial training population $\gP$ to identify conventions.  For a finite population of $m$ policies, the cross-play matrix $\mX \in \R^{m \times m}$ is defined such that $\emX_{ij} = J(\pi_i, \pi_j)$ for each $i,j \in \{1,\ldots,m\}$.

\section{Discovering Conventions}
\label{sec:discovering_conventions}

To simplify the task of discovering a system of conventions over the full policy space $\Pi$ (Section~\ref{sec:conventions}), we focus on partitioning a known, finite population of policies into disjoint conventions.  We implement two algorithms for convention discovery that compute partitionings based on the cross-play matrix induced by these populations.  



\subsection{Partitioning into Conventions}
\label{subsec:partitioning}

Given a finite population $\gP$ of $m$ policies, we first compute a cross-play matrix $\mX$, which captures the varying degrees of compatibility between members of the population.  We consider two approaches to identifying an underlying system of conventions $\gC$, both of which attempt to find systems that serve as a compact representation of $\mX$.  The first approach clusters the rows (or equivalently the columns) of $\mX$ into discrete sets which become the conventions.  The second uses a matrix factorization technique to compute a system of conventions that can be used to reconstruct $\mX$.

\paragraph{Spectral Clustering}

A useful system of conventions should allow us to determine appropriate responses to a policy based only on the knowledge of the convention that policy is following.  A natural approach to convention discovery then is to cluster policies that have similar compatibility structures, and treat these clusters as conventions.  We treat each row $\mX_{i,:}$ as a feature vector representing policy $\pi_i$'s interaction profile, and apply spectral clustering \citep{10.5555/2980539.2980649} to partition policies into groups with similar interaction signatures \citep{10.65109/UMCV8852}.  Conventions emerge naturally as clusters of policies that achieve comparable returns with the same partners, such that a single policy should be able to achieve high return when paired with any member of a given cluster.  The clustering approach requires specifying the number of conventions $q$ as a hyperparameter, which can be determined through eigengap analysis, silhouette analysis, or domain knowledge about the task structure.

\paragraph{Matrix Factorization}

We also evaluate the use of low-rank matrix factorization to discover conventions based on $\mX$.  We approximately decompose $\mX$ as $\mX \approx \mZ \mW \mZ^\top$, where $\mZ \in [0,1]^{m \times q}$ records soft membership weights indicating how strongly each policy belongs to each convention, and $\mW = \text{diag}(\vw)$ captures the expected return for each convention when two policies that implement the same convention are paired together.  The reconstructed cross-play matrix is then $\bar{\mX}_{ij} = \sum_s \emZ_{is} \emW_s \emZ_{js}$.  $\mZ$ and $\mW$ are found via gradient descent on the mean squared reconstruction error $\Vert \mX - \bar{\mX}_{ij} \Vert$.  Finally, each policy is assigned to the convention for which it has the highest membership score.  Like spectral clustering, this approach requires specifying the rank $q$ (the number of conventions) as a hyperparameter.  The requirement that $\mW$ be a diagonal matrix encodes two strong assumptions about the target task: 1) a policy's compatibility with one convention is independent of its compatibility with another, and 2) any two agents that follow the same convention will be compatible with one another.  There are tasks in which these assumptions do not hold, for example, tasks where each player must take on a distinct role (and there is no environmental signal to coordinate these role assignments).  Each of our experimental tasks, however, approximately satisfy these assumptions.

\subsection{Convention-Rebalanced Training}
\label{subsec:rebalanced_training}

We leverage the discovered convention system to improve the effectiveness of the generalist policy learned via best-response training, which depends on the distribution of policies within $\gP$. The existing diversity methods discussed in Section~\ref{sec:related_work} train against the empirical distribution of a learned policy population, which may be skewed across convention labels -- common conventions over-represented while sub-optimal or hard-to-discover conventions under-represented. However, human biases over conventions may differ substantially from the biases that emerge in RL-discovered populations.  We posit that a test-time partner is equally likely to follow any convention for the task, and use a discovered convention system to correct this potential imbalance in the training population. Given a population $\gP$ and its convention partition $\{C_1, \ldots, C_q\}$, we replace sampling $\pi \sim \gP$ with a rebalanced procedure: first sample a convention $C_i$ uniformly from $\{C_1, \ldots, C_q\}$, then sample a policy uniformly at random, with replacement, from $C_i$. This gives every convention equal weight in the training distribution regardless of its size or frequency in $\gP$. A best-response agent is then trained against the rebalanced population $\gP'$ rather than $\gP$.  Experimental results on convention-rebalanced training are discussed in Section~\ref{sec:results_and_discussion}.

\section{ConventionPlay}
\label{sec:conventionplay}
\begin{figure}[t]
    \centering
    \includegraphics[width=0.75\columnwidth]{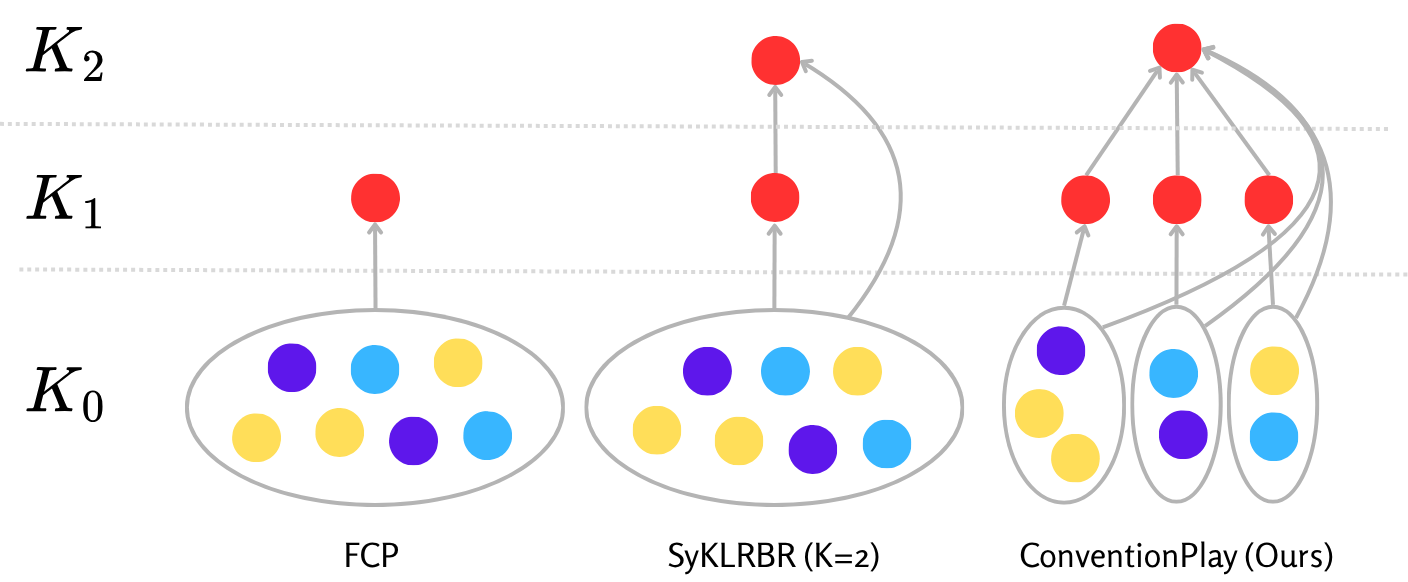}
    \caption{Comparison of the \textit{ConventionPlay} pipeline with existing ad-hoc collaboration methods. Each circle represents an agent. Colors at the $K_0$ level indicate the convention to which each agent adheres. Adaptive agents are shown in red with an arrow from their training population. While baseline approaches diversify the $K_0$ population, we introduce diversity among $K_1$ agents to force the $K_2$ agent to infer its partner's limited capability set.
    }
    \label{fig:convplay_diagram}
\end{figure}

Different approaches to training policies for ad-hoc collaboration can be distinguished by the assumptions they make about the target population of potential partners that they will eventually need to interact with. Convention-rebalanced training (\ref{subsec:rebalanced_training}) models our assumption that a partner may hold any single convention with equal likelihood, but still assumes each partner is limited to just one convention. We now consider that different partners will have varying degrees of ability in coordinating across conventions.  The population model of \textit{ConventionPlay} is described using a variant of the Cognitive Hierarchy (CH) \citep{Camerer2003} model.  Informally, policies at level $k \geq 1$ of the CH model are best-responses to a mixture over policies at the previous levels $0, \ldots, k-1$.  \textit{ConventionPlay} assumes that members of the target population will sit at levels 0 or 1 of the hierarchy, and that the final policy it outputs sits at level 2.  We denote a level-$k$ policy as $\pi^{(k)}$, and construct populations at each level iteratively.  We describe the assumptions made at each level of the hierarchy below, and highlight how the assumptions made by \textit{ConventionPlay} differ from those made by the FCP~\citep{strouseFCPCollaboratingHumans2021} and SyKLRBR~\citep{cuiSyKLRBRKlevelReasoning2021} baselines; this is illustrated in Figure~\ref{fig:convplay_diagram}.

\paragraph{Level-0 ($K_0$): Fixed Convention Agents.}
We would expect a shared convention to emerge among a single population of agents that routinely interact with one another. These agents are brittle: they expect a partner to match their specific mode of coordination. Multiple isolated populations can in turn converge on distinct conventions. The $K_0$ population, $\gP^{(0)}$, represents this aggregate: diverse as a population, but composed of individually fixed-convention agents. Population diversity methods such as MEP, LIPO, and CoMeDi generate and train against a $K_0$ population alone.  In our implementation of \textit{ConventionPlay} the members of the $K_0$ population are trained using PPO in self-play.

\paragraph{Level-1 ($K_1$): Adaptive Agents with Limited Capabilities.}
\label{sec:k1_agents}
Training only against a $K_0$ population ignores the possibility that some partners will be able to adapt to multiple conventions.  A solution is to incorporate agents at the $K_1$ level of the hierarchy into the training population.  SyKLRBR, for example, trains against a $K_1$ population that includes best-responses to the uniform distribution over the full $K_0$ population $\gP^{(0)}$.  Realistic partners, however, may have varying capabilities: they may adapt to multiple conventions, but not all.  \textit{ConventionPlay} captures this possibility by assuming that each $K_1$ agent only encounters a restricted subset of the possible conventions for the task.  For each $K_1$ agent $\pi_i^{(1)} \in \gP^{(1)}$, we define a \textit{capability set} $\gK_i \subset \gC$ as the non-empty subset of conventions agent $i$ supports.  The training distribution for $\pi_i^{(1)}$ is constructed by sampling a training set of policies $\gP_i$ from the conventions in its capability set $\gK_i$:
\begin{equation}
    \gP_i = \bigcup_{C \in \gK_i} \text{Sample}(C, m),
\end{equation}
where $\text{Sample}(C, m)$ draws $m$ policies from convention $C$. The agent is then defined as the best-response to its specific training population: $\pi_i^{(1)} \in \text{BR}(\gP_i)$.  We create a population of $K_1$ agents, $\gP^{(1)}$, by randomly sampling capability sets of size $1 < |\gK_i| < |\gC|$, ensuring coverage of limited-capability partners while maintaining the constraint that no agent supports all conventions.

\paragraph{Level-2 ($K_2$): Robust Ad-hoc Collaborators.}
The $K_2$ agent, $\pi^{(2)}$, is trained to be an optimal collaborator by setting its training population $\gP^{(2)}$ to be the union of the fixed-convention $K_0$ partners and adaptive $K_1$ partners: $\gP^{(2)} = \gP^{(0)} \cup \gP^{(1)}$.  This objective incentivizes $\pi^{(2)}$ to adopt a \textit{probing-and-steering} strategy; it must interact with a partner to infer whether they are a restricted $K_0$ agent, requiring strict adherence to a fixed convention, or an adaptive $K_1$ agent, allowing the $K_2$ agent to steer the team toward the highest-value convention within this partner's specific capability set; the full training procedure is summarized in Algorithm~\ref{alg:conventionplay} (Appendix~\ref{sec:convplay_impl}).

\section{Experimental Setup}

This section introduces the environments and evaluation protocol used to train and test our ad-hoc agents; further details and visualizations for all environments are provided in Appendix~\ref{sec:env_details}.

\subsection{Environments}

We evaluate our approach on multiple versions of three multi-agent environments where coordination on shared conventions is critical. To test robustness across different payoff structures, we configure three variants of each environment: uniform payoffs where all conventions yield equal returns (\emph{non-diff}), partial differentiation where some conventions dominate others (\emph{partial}), and full differentiation where each convention has a unique value (\emph{full}).

\begin{figure}[h]
    \centering
    \includegraphics[width=0.3\columnwidth]{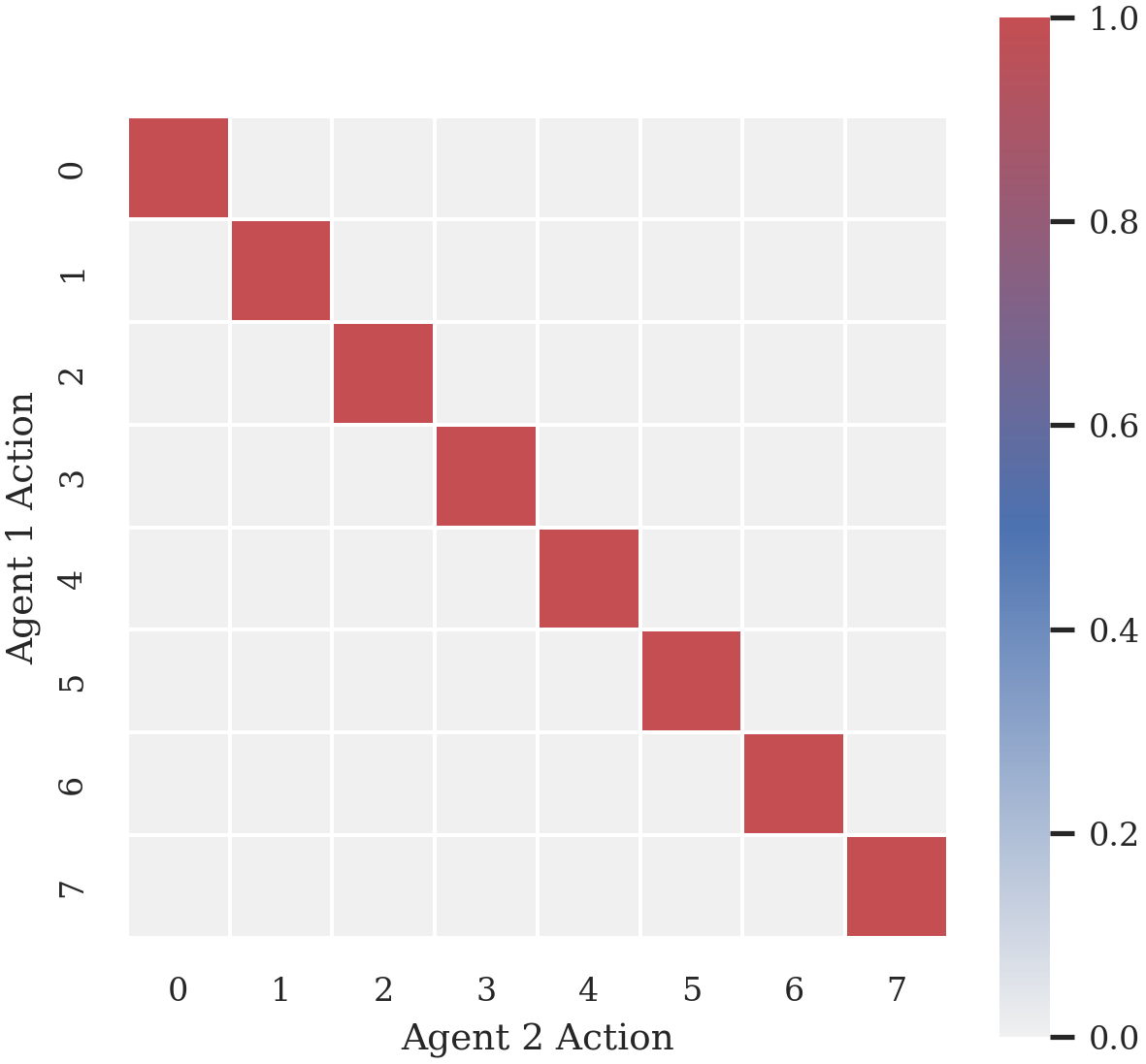}
    \hfill
    \includegraphics[width=0.3\columnwidth]{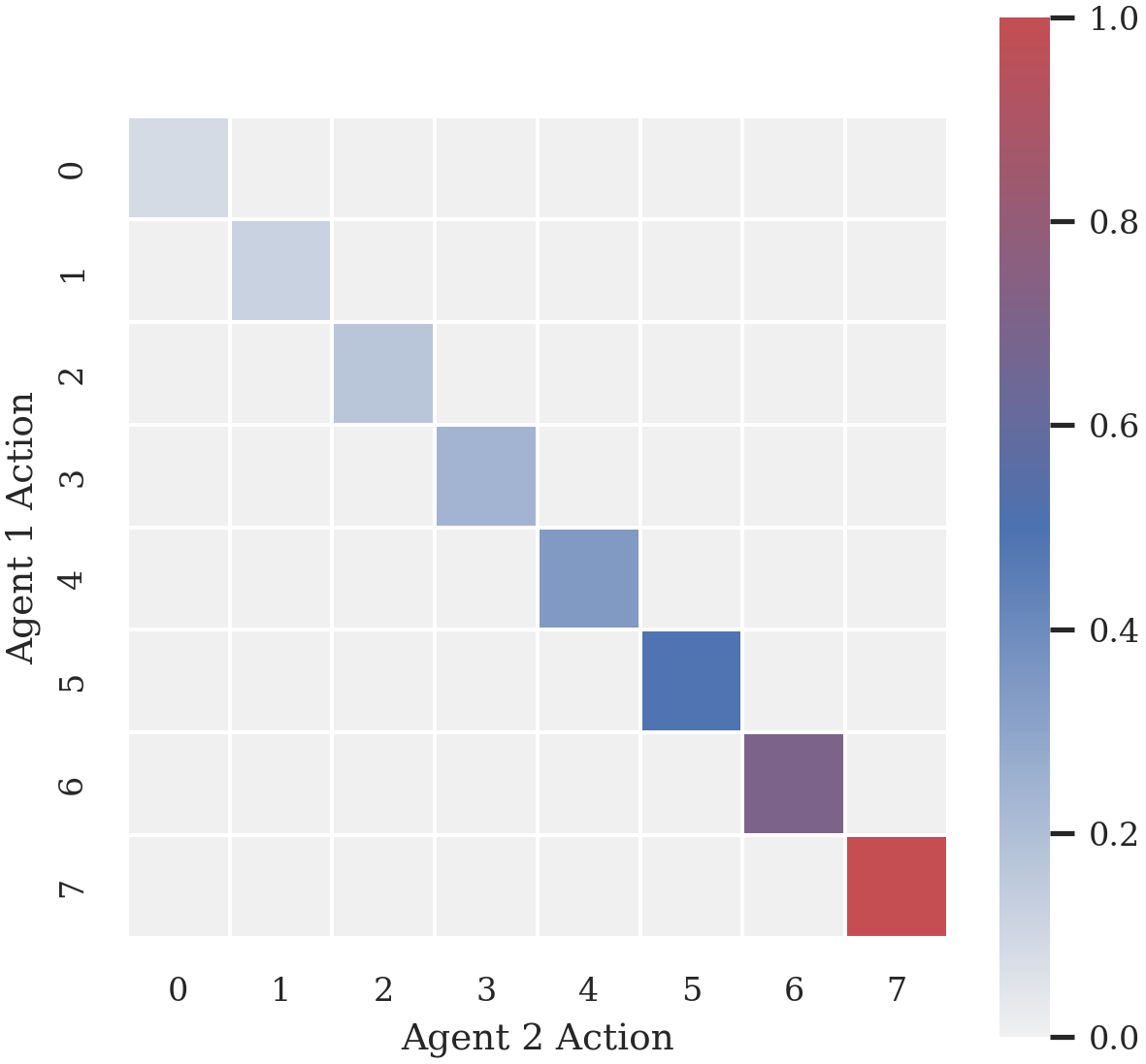}
    \hfill
    \includegraphics[width=0.3\columnwidth]{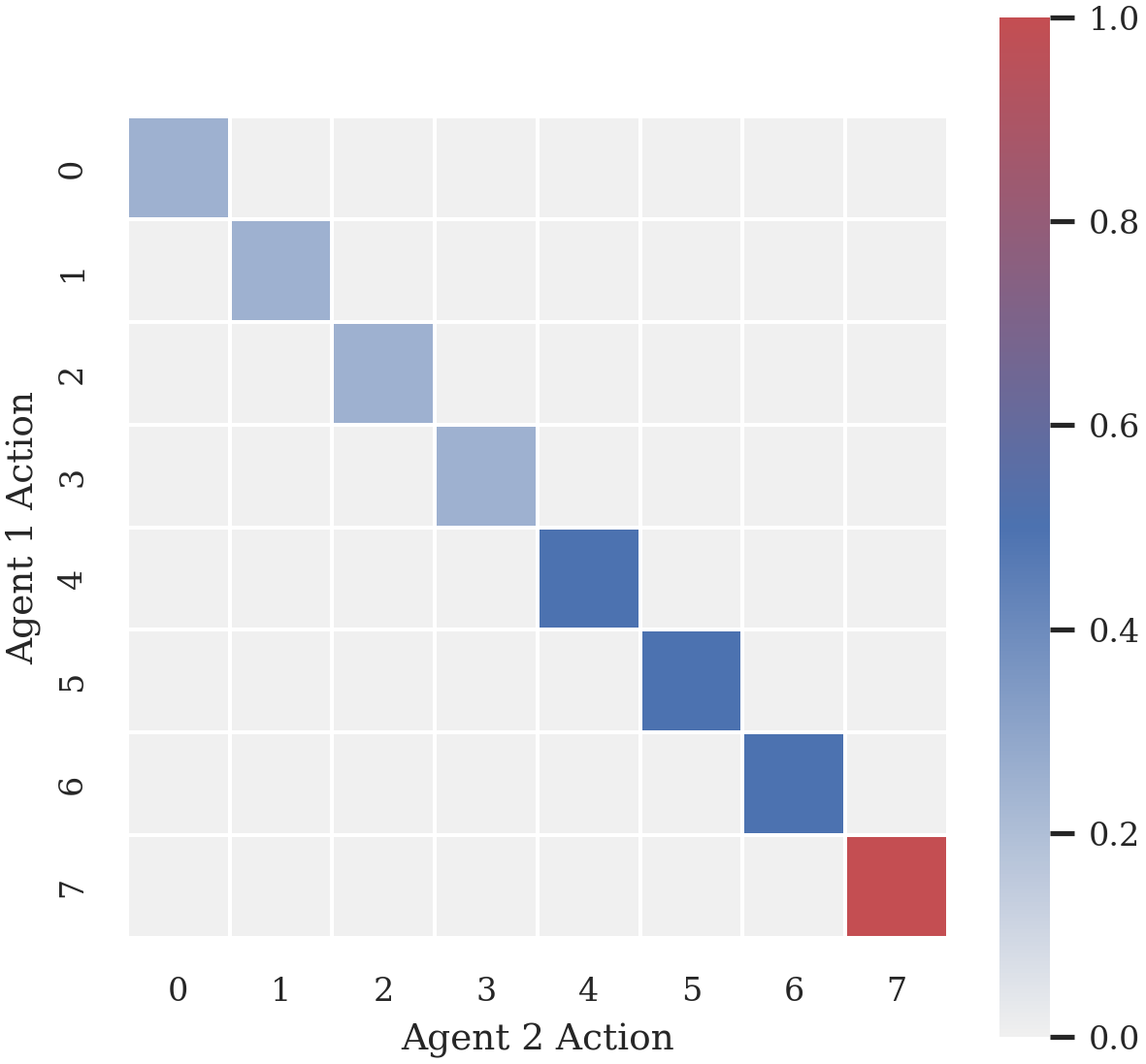}
    \caption{Payoff matrices for the Repeated Matrix Game. Left: Uniform payoffs across conventions. Middle: Partially differentiated payoffs across conventions. Right: Fully differentiated payoffs.}
    \label{fig:matrix_payoffs}
\end{figure}

\paragraph{Repeated Matrix Game.} A coordination task where agents must select matching actions across twenty trials. Agents observe a limited three step action history of their partners. Figure~\ref{fig:matrix_payoffs} shows payoff matrices for uniform and fully differentiated variants. Here, the expected convention is a single coordinated action.

\paragraph{Point Mass Rendezvous (PMR).} A spatial coordination task \citep{charakorn2023generating} where agents must rendezvous at one of six pre-configured landmarks. Agents observe their position, velocity, and distance to landmarks, along with their partner's velocity. The reward signal is sparse, forcing agents to commit to rendezvous at a specific landmark before observing rewards for it. Here, the expected convention is the landmark agents rendezvous at. Like the matrix game, we vary rewards issued at each landmark to create uniform and differentiated convention structures.

\paragraph{Task-Relay.} A novel two-player cooperation game modeling an industrial manufacturing task in which two agents build an object by switching to and applying the appropriate tools. Optimal policies partition labor between the agents so that idling time is minimized, effectively simulating a resource allocation problem. The object's evolution is defined by a task graph with five goal states, each representing a distinct way of manufacturing the object given a set of tools, and each yielding a unique cumulative reward. Here, the expected convention is a choice of which sequence of tool applications to pursue.  Further details for the Task-Relay environment are provided in Appendix~\ref{sec:env_task_relay_details}.

\subsection{Evaluation Protocol}

Our evaluation protocol is designed to probe ad-hoc teamwork capabilities across every level of the cognitive hierarchy. For each environment, we construct three evaluation populations. First, $\gP^{(0)}_{\text{test}}$ uniformly covers the task's known conventions leveraging domain knowledge. Second, $\gP^{(1)}_{\text{sampled}}$ applies the capability-set sampling procedure of Section~\ref{sec:k1_agents} directly to $\gP^{(0)}_{\text{test}}$, using its known ground-truth convention labels in place of a discovered partition. Third, $\gP^{(1)}_{\text{heur}}$ consists of hand-coded heuristic agents that implement limited-capability behavior directly -- for example, a stubborn tit-for-tat agent in the repeated matrix game that matches its partner's action only when that action falls within a specific capability set. Environment-specific details for developing $\gP^{(0)}_{\text{test}}$ and $\gP^{(1)}_{\text{heur}}$ are provided in Appendix~\ref{sec:env_details}. Because $\gP^{(0)}_{\text{test}}$, $\gP^{(1)}_{\text{sampled}}$, and $\gP^{(1)}_{\text{heur}}$ are constructed independently of any training population, they may include behaviors or behavioral noise that no evaluated $K_2$ method encountered during training; performance against these populations reflects out-of-distribution generalization.

We compare against \textbf{FCP} \citep{strouseFCPCollaboratingHumans2021}, which introduces behavioral diversity through intermediate checkpoints, and \textbf{SyKLRBR} \citep{cuiSyKLRBRKlevelReasoning2021}, representing state-of-the-art cognitive hierarchy modeling for zero-shot coordination; for structural parity, we evaluate a two-level SyKLRBR hierarchy grounded in our learned $K_0$ population in addition to the canonical five-level hierarchy. We also include the population-diversity methods \textbf{MEP}, \textbf{LIPO}, and \textbf{CoMeDi}. For CoMeDi, we evaluate both the original implementation (CoMeDi (BC)) and a variant, CoMeDi (BR), that replaces its joint self-play and imitation objective with a single best-response agent trained on our recurrent architecture. Appendix~\ref{sec:baselines} details the motivation for all baselines. 

\section{Results and Discussion}
\label{sec:results_and_discussion}

We first validate convention discovery in isolation, through an experiment in convention-rebalancing. We then evaluate our generalist $K_2$ agent. For each experiment we test convention discovery with both spectral clustering (Spectral) and matrix factorization (MatFact). All results are reported in terms of normalized reward and coordination efficiency, defined respectively in Appendices~\ref{sec:normalized_return} and~\ref{sec:coord_efficiency}. Hyperparameters and implementation details for all experimental runs are reported in Appendices~\ref{sec:hyperparameters} and \ref{sec:convplay_impl}. We additionally study the effect of $K_1$ population size on \textit{ConventionPlay} in Appendix~\ref{sec:k1_size_effect}.

\paragraph{Convention Rebalancing}
\label{subsec:rebalancing_results}

We validate the convention identification and rebalancing of Section~\ref{sec:discovering_conventions} using a variant of the matrix game in which unequal action-set sizes across conventions cause $K_0$ training to naturally yield a heavily imbalanced population. Figure~\ref{fig:rebalance_bars} compares a best-response agent trained on this imbalanced population against ones trained on rebalanced populations $\gP'$, evaluated on a $K_0$ test population that is uniform over conventions. Rebalancing improves normalized return by a margin of up to 40\%, confirming that correcting for skewed convention representation during training improves generalization to an unbiased test-time partner distribution. Full experimental setup and additional figures are provided in Appendix~\ref{sec:rebalancing}.  

\paragraph{Performance on Test Populations}
\KZeroRewardsTable
\KOneRewardsTable

\begin{figure}[h]
    \centering
    \includegraphics[width=\textwidth]{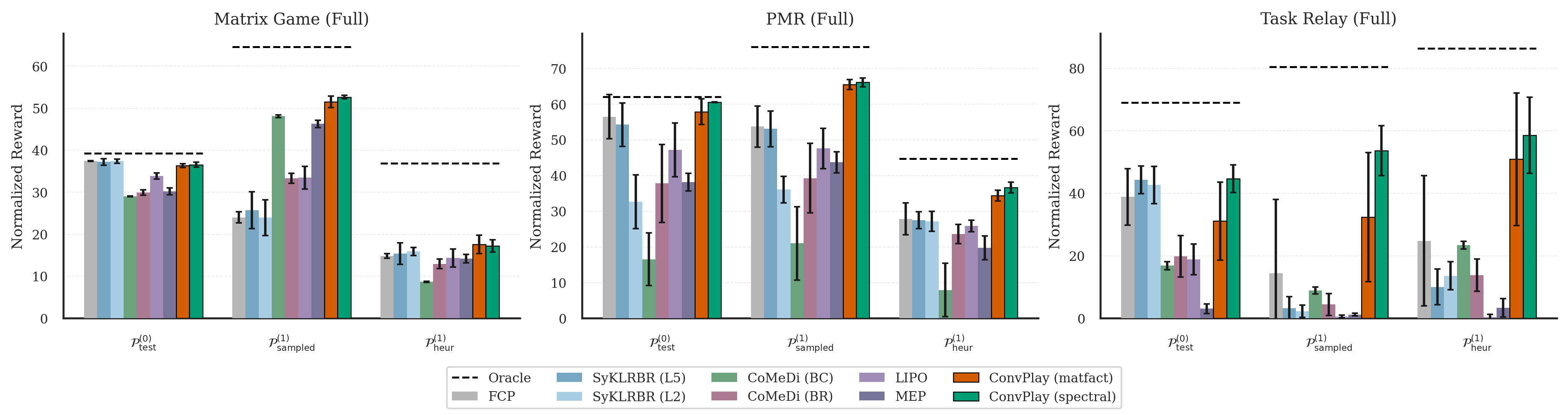}
    \caption{Normalized reward under full payoff differentiation across all three environments (Matrix Game, PMR, Task-Relay) and all three evaluation populations ($\gP^{(0)}_{\text{test}}$, $\gP^{(1)}_{\text{sampled}}$, $\gP^{(1)}_{\text{heur}}$). The same pattern can be observed in all settings: \textit{ConventionPlay} shows a modest cost against fixed-convention partners, but outperforms baselines against various types of adaptive partners.}
    \label{fig:bar_reward_full}
\end{figure}

The results across all three evaluation populations and environments---Tables~\ref{tab:k0_rewards}, \ref{tab:k1_rewards}, \ref{tab:k1_heur_rewards}---follow a similar shape: parity with baselines when conventions are non-differentiated, and both a 10--50\% gain against adaptive partners and a modest 0--5\% reduction against fixed-convention partners when conventions are differentiated (Figure~\ref{fig:bar_reward_full}). Non-differentiated settings pose little challenge, since convergence to any shared convention is equally good. The tradeoff in the differentiated setting follows directly from the probing-and-steering behavior induced by our training population: baseline agents lack any mechanism for distinguishing a partner capable of a higher-value convention from one that is not, and so commit to the first shared convention they discover, even when a superior alternative lies within the partner's capability set. \textit{ConventionPlay}, by contrast, first probes for a higher-value convention before committing---a strategy that pays off against adaptive partners, and costs only a little against rigid ones, since the agent cannot immediately tell the two apart. This performance gap is most pronounced in Task-Relay, where baselines perform particularly poorly against $\gP^{(1)}_{\text{sampled}}$ and $\gP^{(1)}_{\text{heur}}$ partners. This stems from Task-Relay's susceptibility to a co-adaptation deadlock: each agent repeats non-committal actions, waiting for the other to commit first---precisely the dynamic \textit{ConventionPlay} is designed to break.

\begin{figure}[h]
    \centering
    \includegraphics[width=0.9\columnwidth]{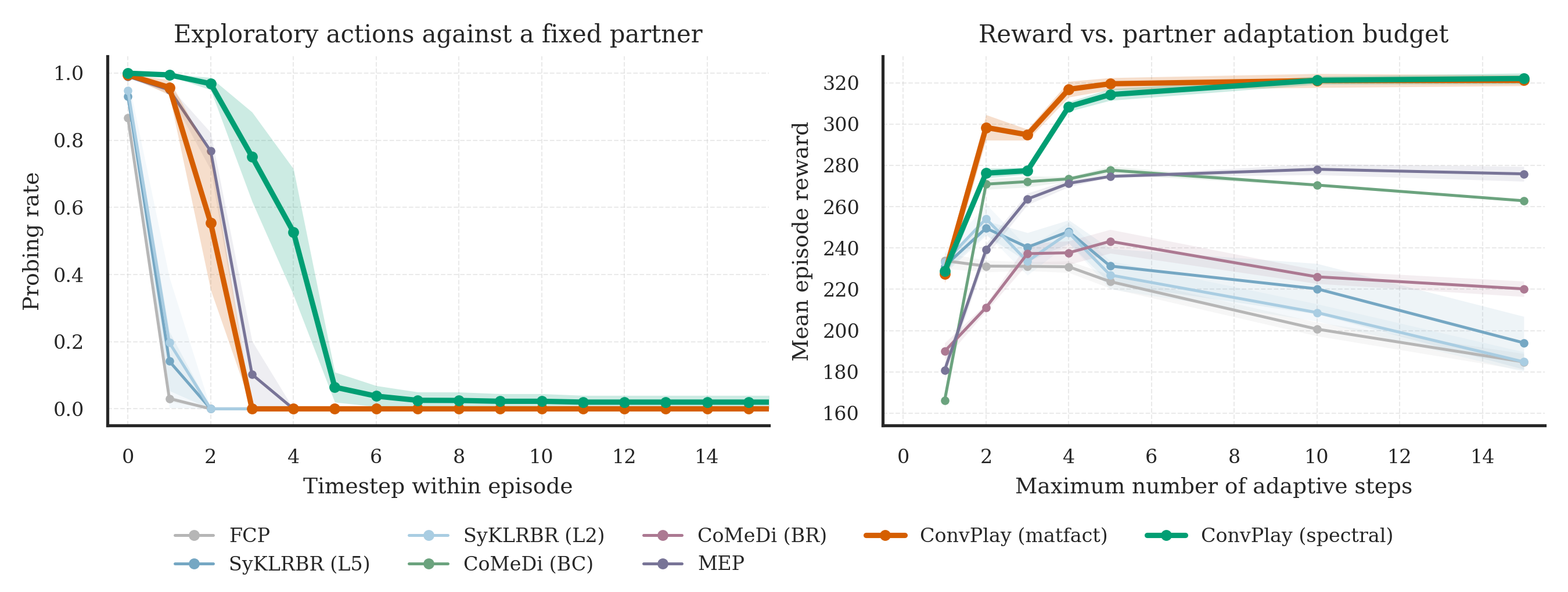}
    \caption{Left: fraction of various methods applying a probing (i.e., uncoordinated) action to a fixed-convention partner; \textit{ConventionPlay} can be seen to probe for significantly longer, explaining its reduced performance with the $\gP^{(0)}$ population. Right: performance on a modified $\gP^{(1)}_{\text{sampled}}$ in which partners adapt for only a fixed number of steps before committing, tracing \textit{ConventionPlay}'s advantage across a spectrum of partner adaptivity.}
    \label{fig:action_signaling}
\end{figure}

\paragraph{Analysis of Probing}
We examine the emergent probing behavior by analyzing trajectories from interactions in the Matrix Game under fully differentiated payoffs.  The left panel of Figure~\ref{fig:action_signaling} shows each method's interaction with a fixed-convention partner committed to the lowest-value convention -- a scenario in which the ad-hoc agent is most likely to ``probe'': deviate from its partner's current action in an attempt to induce the partner to adapt toward a higher-value convention. Whereas most baseline methods converge to the partner's convention within 1--3 time steps, \textit{ConventionPlay} probes for 3--6 time steps before converging, explaining its modest reward cost against $\gP^{(0)}_{\text{test}}$.  The right panel evaluates ad-hoc methods coordinating with a modified version of the learned policies in $\gP^{(1)}_{\text{sampled}}$ in which partner policies are run for a fixed number of steps before committing to their last action, allowing us to trace \textit{ConventionPlay}'s performance across a spectrum of partner adaptivity. At low adaptivity, the agent has little time to gather information about its partner and must coordinate based on what it has already observed; this setting effectively reduces the partner to a $K_0$ agent, and \textit{ConventionPlay} matches the expected 0--5\% performance drop compared to leading baselines. As adaptivity increases, however, \textit{ConventionPlay} increasingly outperforms all baselines by effectively inferring the partner's hidden capabilities.  A similar probing behavior can be visualized with individual PMR trajectories and is shown in Appendix~\ref{sec:pmr_probing}. 

\section{Conclusion and Future Work}
\label{sec:conclusion}

In this work, we have introduced \textit{ConventionPlay}, a reinforcement learning-based approach for ad-hoc collaboration that moves beyond purely reactive partner adaptation. By introducing strategic diversity into the $K_1$ layer of a cognitive hierarchy, agents learn to dynamically balance leading and following. Our experimental results demonstrate that when partners themselves are capable of adapting across multiple conventions, \textit{ConventionPlay} outperforms existing methods by actively probing a partner's capabilities and steering the team toward high-value conventions.

\paragraph{Limitations}
A primary limitation in our approach is the reliance on cross-play returns for convention identification. Incorporating other behavioral signals---state-action trajectories or policy similarity \citep{lupuTrajectoryDiversityZeroShot2021}---could provide more nuanced representations of agent capabilities, which would in turn improve the construction of capability-limited $K_1$ training sets. Additionally, our evaluation focuses on relatively small populations of conventions; scaling to environments with many overlapping or complex convention structures remains an open question.

\subsection*{AI use statement}

We used AI in a number of targeted ways, and all AI-assisted work was heavily reviewed and revised by the authors.

For software development, we used AI to build a lightweight harness for parallelizing large experimental runs with a declared dependency graph, to vectorize the Task-Relay environment, and to write utilities converting experiment artifacts into formatted data and visualizations (e.g. Table~\ref{tab:k0_rewards}, Figures~\ref{fig:task_relay_combined} and \ref{fig:pmr_probe}). Prompts were narrowly scoped, producing small diffs that we carefully reviewed. 

When developing experiments, we used AI to help debug results and suggest hyperparameters, primarily to maximize throughput on available compute (e.g., data collections rate and batch sizes).

We used AI for grammar and language refinement while writing this document.

\bibliography{all_items}
\bibliographystyle{iclr2027_conference}

\appendix
\section{Environments}
\label{sec:env_details}

This section describes environments used for evaluation and specifically details how domain knowledge is leveraged to construct $\gP^{(0)}_{\text{test}}$ and $\gP^{(1)}_{\text{heur}}$ populations.

\subsection{Repeated Matrix Game}
The repeated matrix game requires agents to coordinate on matching actions across 20 trials. Figure~\ref{fig:matrix_payoffs} shows payoff matrices for the uniform and fully differentiated variants used in our experiments.

\textbf{Observation space.} Each agent observes its partner's three most recent actions.

\textbf{Action space.} Each agent selects one of $N$ discrete actions per trial; in our configured variant, $N=8$.

\textbf{Evaluation agents.} For the $\gP^{(0)}_{\text{test}}$ set, we create hard-coded policies that each play a single fixed action throughout the episode. For the $\gP^{(1)}_{\text{heur}}$ set, we implement a stubborn tit-for-tat agent that plays a fixed default action until it observes its partner playing an action within its own capability set, at which point it switches to mirroring that action for the remainder of the episode. We instantiate many such agents, parameterized by a pair $(a_0, a_1)$ of default and candidate actions from $\{1,\ldots,N\}$ -- e.g., $(0,1)$, $(0,2)$, $(1,3)$, and so on -- covering a range of default/candidate combinations.

\subsection{Point Mass Rendezvous (PMR)}
Point Mass Rendezvous (PMR) is a spatial coordination task \citep{charakorn2023generating} implemented using VMAS \citep{bettini2022vmas}, in which agents must rendezvous at one of 6 pre-configured landmarks. 

\textbf{Observation space.} Each agent observes its own position and velocity, its distance to each landmark, and its partner's velocity.

\textbf{Action space.} Each agent outputs a continuous 2D action controlling its planar acceleration.

\textbf{Evaluation agents.} For the $\gP^{(0)}_{\text{test}}$ set, we create policies that deterministically drive to each one of the pre-configured landmarks. For the $\gP^{(1)}_{\text{heur}}$ set, we implement a stubborn tit-for-tat agent that drives toward a default landmark for a fixed initial period before potentially switching to a landmark in its capability set if the partner is closest to it. We instantiate many such agents, parameterized by a pair $(l_0, l_1)$ of default and candidate landmarks -- e.g., $(0,1)$, $(0,2)$, $(1,3)$, and so on -- covering a broad range of default/candidate combinations.

\subsection{Task-Relay}
\label{sec:env_task_relay_details}

\begin{figure}[h]
    \centering
    \begin{subfigure}[b]{0.8\columnwidth}
        \centering
        \includegraphics[width=\textwidth]{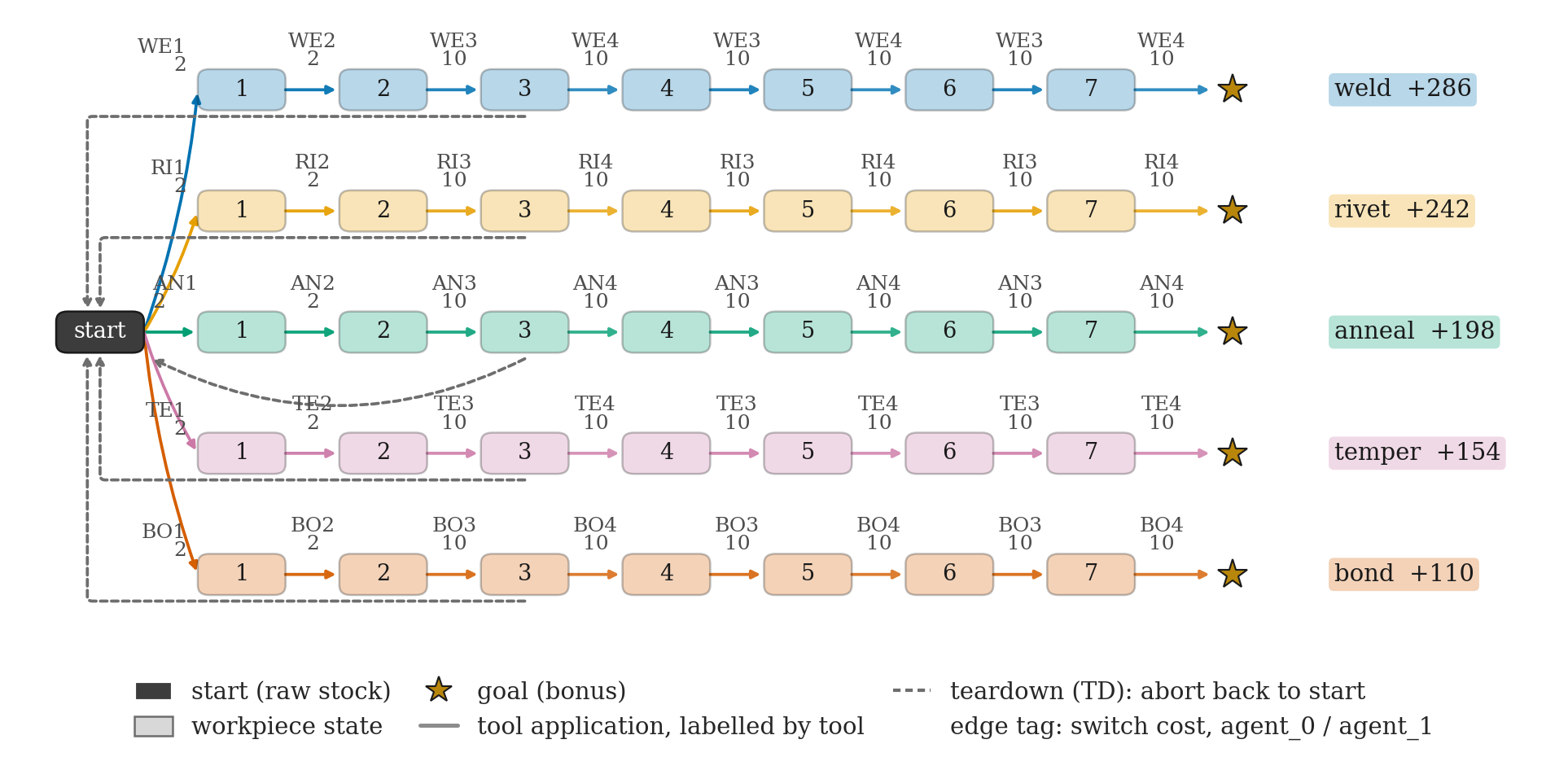}
        \caption{The task graph for constructing an object underlying Task-Relay. The dark node is the shared start (raw stock) state; each colored row is a distinct strategy, advancing through a chain of intermediate workpiece states via applications of its associated tool, and terminating in a goal state (star) labeled with its completion bonus and tool switch cost. Dashed arrows denote a teardown action that aborts a strategy back to the start state.}
        \label{fig:task_relay_graph}
    \end{subfigure}
    \vspace{1em}
    \begin{subfigure}[b]{0.9\columnwidth}
        \centering
        \includegraphics[width=\textwidth]{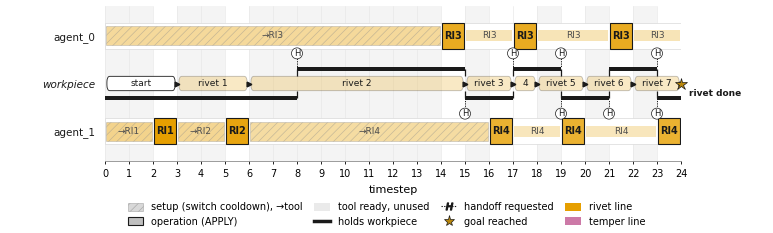}
        \caption{Example resolved trajectory for the \textit{rivet} strategy, showing each agent's tool timeline and the workpiece's progression over the episode.}
        \label{fig:tr_fab_traj}
    \end{subfigure}
    \caption{Task-Relay task structure and an example resolved trajectory.}
    \label{fig:task_relay_combined}
\end{figure}

Task-Relay requires two agents to jointly build a single object by applying tools that advance the object along a directed graph of object states; only one agent may hold the object at a time, and only the holder may apply a tool. Figure~\ref{fig:task_relay_graph} shows the task graph used in our experiments; the graph has five manufacturing paths (weld, rivet, anneal, temper, bond) branching from a shared start state. Switching tools is not instantaneous, so the two agents must interleave staging the next tool and holding the object to avoid wasted time. 

\textbf{Observation space.} Each agent observes its own and its partner's equipped tool, in-progress switch target, and remaining switch cooldown; whether it holds the object; a one-hot encoding of the object's current node; which tools have a live edge at the
current node and one hop ahead; and the fraction of the episode elapsed.

\textbf{Action space.} Each agent selects one of NOOP, APPLY, HANDOFF, or SWITCH to one of $K$ tools per step; in our configured variant, $K=21$.

\textbf{Evaluation agents.} For the $\gP^{(0)}_{\text{test}}$ set, we train self-play populations under environment variants, where each variant biases self-play toward specializing on a possible manufacturing path. For the $\gP^{(1)}_{\text{heur}}$ set, we implement a hard-coded policy that never initiates or contests the line choice made by its partner, but will be a competent partner if the chosen path is within the agent's repertoire. We instantiate both singleton (e.g. $\{\text{weld}\}$) and pair (e.g. $\{\text{weld}, \text{rivet}\}$) repertoires, so scoring optimally requires identifying which line(s) are supported by the partner.

\section{Baselines and Hyperparameters}
\label{sec:hyperparameters}

\subsection{Choice of Baselines}
\label{sec:baselines}

We compare \textit{ConventionPlay} against five baselines: FCP, SyKLRBR, MEP, LIPO, and CoMeDi.

\textbf{FCP} \citep{strouseFCPCollaboratingHumans2021} is a standard and widely used baseline for generalist ad-hoc coordination.

\textbf{SyKLRBR} \citep{cuiSyKLRBRKlevelReasoning2021} is the closest existing baseline to our approach, as it is grounded in cognitive hierarchy reasoning; comparing against it isolates the specific contribution of \textit{ConventionPlay}'s capability-limited $K_1$ population relative to a hierarchy that does not model partner capability limits explicitly.

\textbf{MEP} \citep{zhaoMEPMaximumEntropy2022}, \textbf{LIPO} \citep{charakorn2023generating}, and \textbf{CoMeDi} \citep{sarkarComediDiverseConventions2023} each contribute a distinct method for generating diversity within the $K_0$ population and represent more recent points of comparison. For CoMeDi specifically, we evaluate two versions. The published version, \textbf{CoMeDi (BC)}, trains the ad-hoc agent with a self-play plus behavior-cloning objective that clones the CoMeDi population; this design targets human-AI collaboration settings, where the partner is assumed to abruptly switch strategies mid-episode, and behavior cloning helps the agent track that shift. We additionally construct \textbf{CoMeDi (BR)}: a variant that uses the CoMeDi population purely as a source of $K_0$ diversity, and trains a single best-response agent against it with the same recurrent architecture used for our other baselines and \textit{ConventionPlay} itself; this allows us to disentangle CoMeDi's contribution to population diversity from its approach to leveraging that diversity for ad-hoc collaboration. 

\subsection{Self-Play and Best-Response Hyperparameters}

Self-play policies are trained with DQN \citep{tampuu2017multiagent} and PPO \citep{schulman2017proximal}; training uses feedforward networks. All best-response policies are trained with PPO and use recurrent (LSTM) policy networks to handle the partial observability of inferring partner conventions. Additionally for all best-response training, the value networks include an auxiliary head that predicts the partner identity to aid partner identification. All policies were trained using 5 random seeds. For best-response training, 

Table~\ref{tab:selfplay-hyperparams} reports hyperparameters for self-play ($K_0$) training, and Table~\ref{tab:adaptiveplay-hyperparams} reports hyperparameters for best-response ($K_1$/$K_2$) training, by environment.

\SelfplayHyperparamsTable
\AdaptiveplayHyperparamsTable

\subsection{Baseline Population Parameters}
\label{sec:baseline_population_params}

Table~\ref{tab:k0-population-params} reports the $K_0$ population size used for each method, by environment. Table~\ref{tab:conventions-per-env} reports the size of the $K_1$ test (evaluation) population, shared across all methods, and the number of conventions $q$ used as a hyperparameter for both convention partitioning approaches of Section~\ref{subsec:convention_partioning}.

\begin{table}[h]
\centering
\caption{$K_0$ population size for each method, by environment.}
\label{tab:k0-population-params}
\begin{tabular}{lccc}
\toprule
Method & Matrix Game & PMR & Task-Relay \\
\midrule
MEP & 20 & 12 & 10 \\
LIPO & 20 & 12 & 5 \\
CoMeDi (BC) / CoMeDi (BR) & 20 & 12 & 10 \\
FCP & 40 & 30 & 30 \\
SyKLRBR & 40 & 30 & 30 \\
\textit{ConventionPlay} & 40 & 30 & 30 \\
\bottomrule
\end{tabular}
\end{table}

\begin{table}[h]
\centering
\caption{$K_1$ test population size and number of conventions $q$, by environment.}
\label{tab:conventions-per-env}
\begin{tabular}{lccc}
\toprule
 & Matrix Game & PMR & Task-Relay \\
\midrule
$K_1$ test population size & 50 & 12 & 10 \\
$q$ (number of conventions) & 8 & 6 & 5 \\
\bottomrule
\end{tabular}
\end{table}

\section{Experiment Implementation Details}
\label{sec:convplay_impl}

Algorithm~\ref{alg:conventionplay} summarizes the full \textit{ConventionPlay} training procedure; the following subsections detail each stage.

\begin{algorithm}
\caption{ConventionPlay}
\label{alg:conventionplay}
\begin{algorithmic}[1]
\State \textbf{Input:} Number of $K_1$ agents $n$
\State $\gP^{(0)} \gets \texttt{TrainK0Population()}$
\Comment{Stage 1: Generate and partition $K_0$ agents}
\State $\{C_1, \ldots, C_q\} \gets \texttt{PartitionIntoConventions}(\gP^{(0)})$
\State $\gP^{(1)} \gets \emptyset$
\For{$j = 1$ \textbf{to} $n$}
\Comment{Stage 2: Train diverse $K_1$ agents}
  \State Sample capability set $\gK_j \subseteq \{C_1, \ldots, C_q\}$ with $1 < |\gK_j| < q$
  \State $\gP_{j} \gets \bigcup_{C \in \gK_j} \texttt{Sample}(C, m)$
  \State $\pi_{j}^{(1)} \gets \text{BR}(\gP_{j})$
  \State $\gP^{(1)} \gets \gP^{(1)} \cup \{\pi_{j}^{(1)}\}$
\EndFor
\State $\pi^{(2)} \gets \text{BR}(\gP^{(0)} \cup \gP^{(1)})$\Comment{Stage 3: Train robust $K_2$ agent}
\State \Return $\pi^{(2)}$
\end{algorithmic}
\end{algorithm}

\subsection{K0 Population Construction}
\label{subsec:base_population}

We generate our $K_0$ population by training independent PPO \citep{schulman2017proximal} across multiple random seeds. While specialized methods like TrajeDI or LIPO could also be used, we found that random initialization provided sufficient behavioral diversity. This approach does not guarantee that the population will partition into the well-structured system of incompatible policies described in our formalism; in some tasks, we may encounter complex, non-transitive compatibility structures. We therefore verify this inter-convention incompatibility empirically to ensure our agents provide a clear foundation of distinct strategies for subsequent training levels.

\subsection{Convention Partitioning}
\label{subsec:convention_partioning}

For spectral clustering, we first normalize the cross-play matrix $\mX$ based on the expected reward range for the environment, yielding $\tilde{\mX}$. We compute an affinity matrix $\mA \in \R^{m \times m}$ using a standard RBF kernel over each pair of agents' normalized feature vectors, $\emA_{ij} = \exp(-\gamma \lVert \tilde{\mX}_{i,:} - \tilde{\mX}_{j,:} \rVert^2)$ with $\gamma = 1.0$. Clustering is then computed via the off-the-shelf implementation from scikit-learn \citep{scikit-learn}.

For matrix factorization, we parameterize $\mZ$ via a learned logit matrix $\mE \in \R^{m \times q}$, with entries passed through a temperature-scaled softmax: $\mZ = \softmax(\mE / \tau)$. We optimize $\mE$ and $\mW$ via gradient descent (Adam) to minimize a mean-squared reconstruction loss between $\mX$ and $\bar{\mX}$. In this work, temperature $\tau$ is annealed from $1.0 \to 0.01$ during training, implementing a relaxation from soft membership assignments (high $\tau$) to near one-hot convention assignments (low $\tau$). This need not be the case, however; an adjusted temperature or alternate activation function could allow downstream applications to take advantage of overlapping or probabilistic convention structures.

For both convention partitioning approaches (spectral clustering, and matrix factorization), we set the expected number of conventions $q$ based on domain knowledge (detailed in Table~\ref{tab:conventions-per-env}, Appendix~\ref{sec:baseline_population_params}). However, this can be extended to a parameter-free approach via silhouette analysis or eigen decomposition of the cross-play matrix.

\subsection{K1 Population Construction}

When constructing the $K_1$ training population, we sample capability sets $\gK_j$ for each of the $n$ agents such that $1 < |\gK_j| < q$. To ensure diversity in partner flexibility while maintaining computational efficiency, we prioritize smaller capability sets during sampling. Specifically, we sample capability sets in order of increasing size: we first sample agents with 2-convention capability sets, then 3-convention sets, and so on, until we have constructed the full $K_1$ population of size $n$. This stratified sampling strategy ensures that the final $K_2$ agent encounters partners with varying degrees of adaptability, from those capable of adapting across just two conventions to those capable of adapting across nearly all conventions.

For each sampled capability set $\gK_j$, we then construct the training population $\gP_{j}$ by uniformly sampling $m$ policies from each convention in $\gK_j$, where $m$ is the number of policies per convention. This ensures that each $K_1$ agent receives balanced exposure to the conventions in its capability set. Since we uniformly sample capability subsets but different subsets may contain different numbers of $K_0$ policies, individual agents may be non-uniformly represented across the full $K_1$ population.

\begin{figure}[h]
    \centering
    \includegraphics[width=0.99\columnwidth]{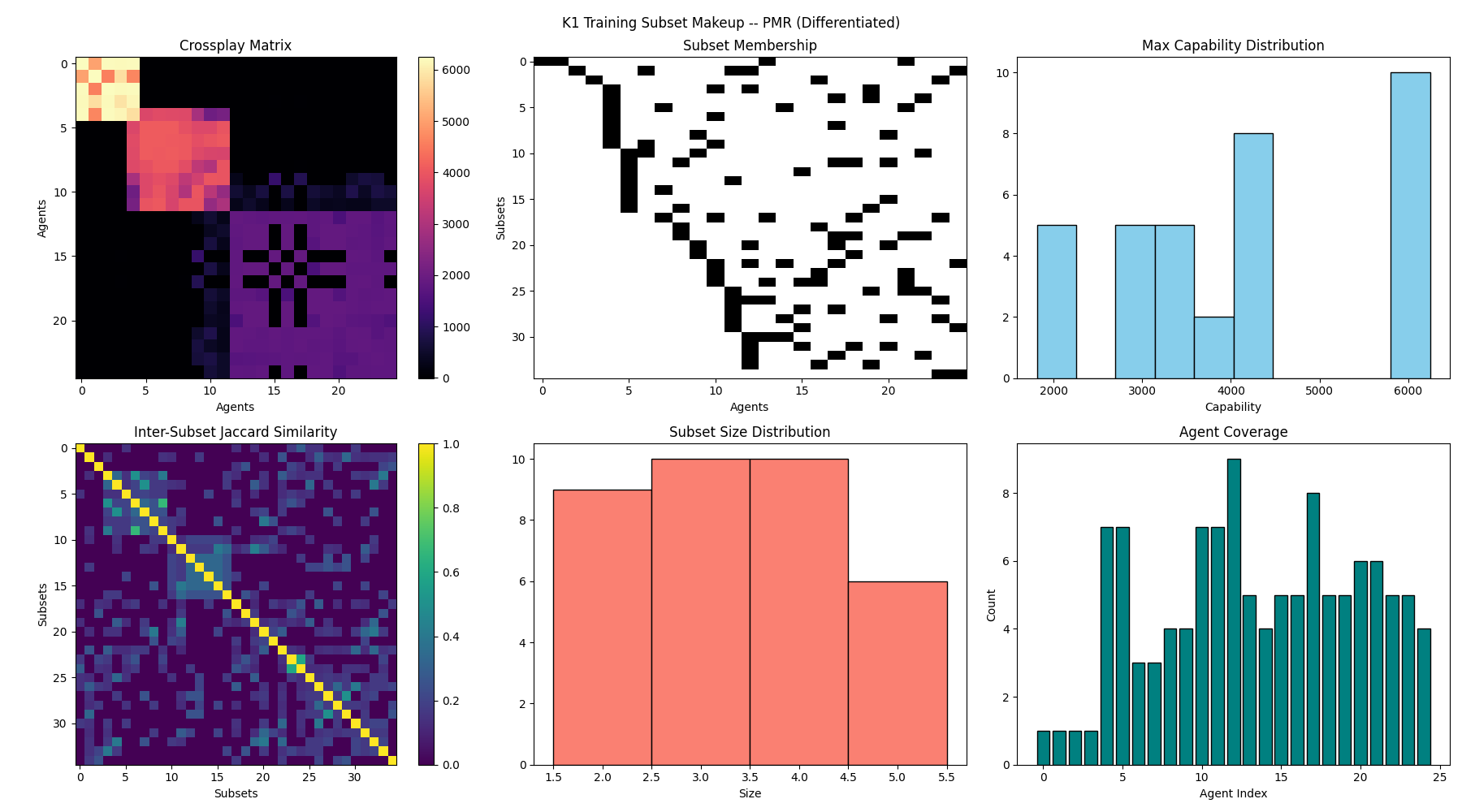}
    \caption{Makeup of the generated $K_1$ subsets. Top left: cross-play matrix of the $K_0$ population, clustered into convention sets. Top middle: sampling of individual policies from each convention to form training distributions for individual $K_1$ agents. Top right: distribution of maximum achievable returns across $K_1$ subsets, reflecting varying capability levels. Bottom left: Jaccard similarity between generated $K_1$ subsets. Bottom right: $K_1$ subset sizes and frequency with which each $K_0$ policy appears across all $K_1$ training distributions.}
    \label{fig:subset_makeup}
\end{figure}

Figure~\ref{fig:subset_makeup} visualizes the composition of generated $K_1$ subsets across multiple panels, illustrating the cross-play structure of the $K_0$ population, stratified sampling across conventions, capability variation among $K_1$ agents, and the distribution of $K_0$ policies across the full $K_1$ population.

\section{Supplementary Experimental Results}
\subsection{Convention-Rebalanced Training}
\label{sec:rebalancing}

This section reports the population-rebalancing experiment introduced in Section~\ref{subsec:rebalancing_results}, isolating the effect of rebalancing a training population across conventions independent of the full \textit{ConventionPlay} hierarchy.

\begin{figure}[h]
    \centering
    \begin{subfigure}[b]{0.47\columnwidth}
        \centering
        \includegraphics[width=0.8\textwidth]{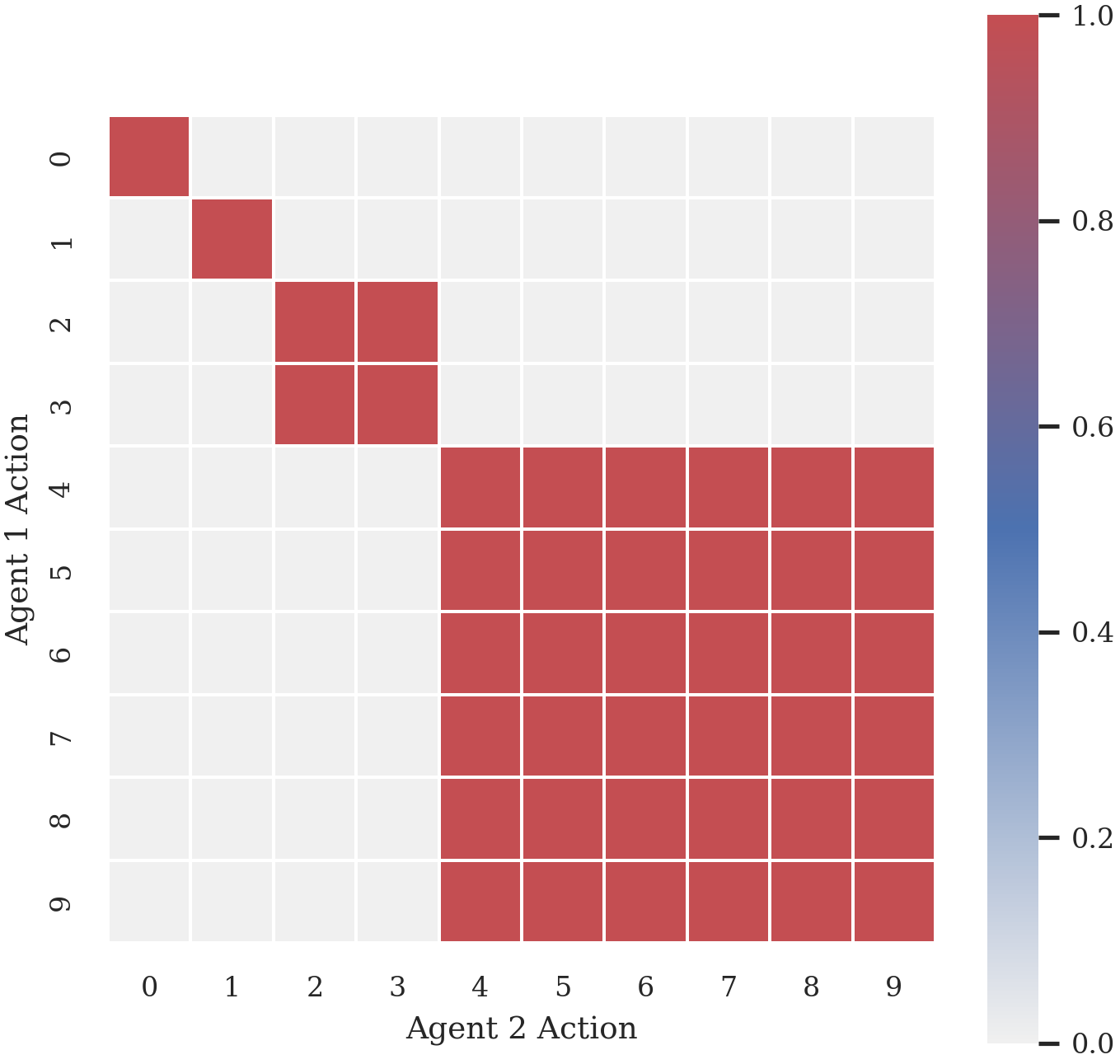}
        \caption{Payoff matrix for the Matrix Game variant: four equal-reward conventions with unequal numbers of compatible actions (1, 1, 2, and 6), following the payoff formulation of \citet{charakorn2023generating}.}
        \label{fig:mx_commonE_payoff}
    \end{subfigure}
    \hfill
    \begin{subfigure}[b]{0.47\columnwidth}
        \centering
        \includegraphics[width=0.8\textwidth]{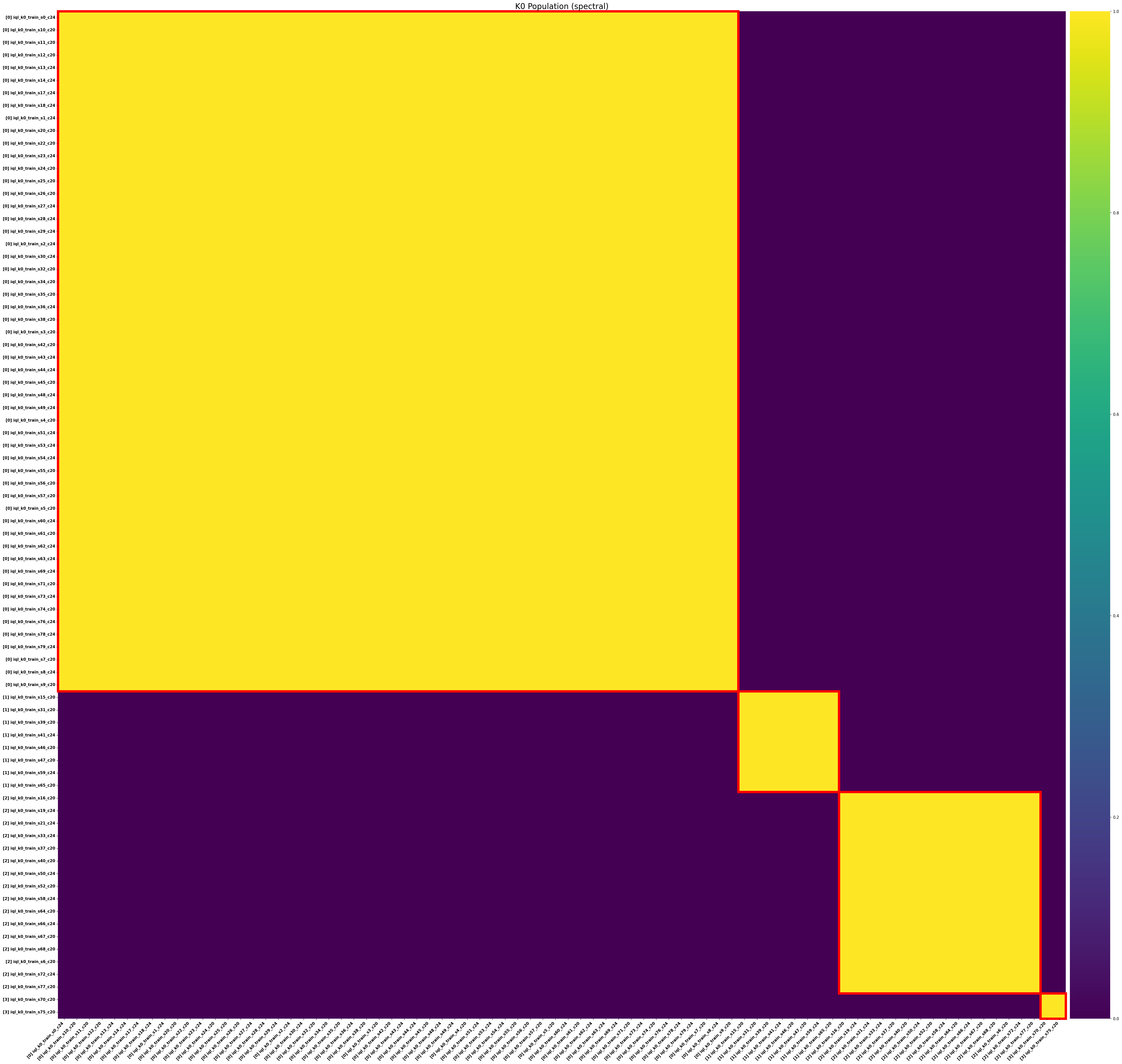}
        \caption{Cross-play matrix of a $K_0$ population trained via self-play on this variant, clustered into its four discovered conventions. The population is heavily skewed toward the largest, most discoverable convention.}
        \label{fig:mx_commonE_unbalanced}
    \end{subfigure}
    \caption{Setup for the convention-rebalancing experiment: a matrix-game variant designed so that some conventions are harder to discover than others, and the resulting imbalance in a self-play $K_0$ population.}
    \label{fig:mx_commonE_setup}
\end{figure}

\paragraph{Setup} We construct a variant of the matrix game whose payoff structure is designed to make some conventions harder to discover than others: every convention here yields the same reward, but differs in how many compatible actions it spans (1, 1, 2, and 6 actions respectively---shown in Figure~\ref{fig:mx_commonE_payoff}). This mirrors the hard-to-find CMG-H setup of \citet{charakorn2023generating}, in which smaller compatible-action sets are less likely to be discovered by self-play under random exploration. As expected, training a $K_0$ population under self-play skews heavily toward easy-to-find conventions (Figure~\ref{fig:mx_commonE_unbalanced}), with the hardest-to-find conventions represented by only a small number of policies. Because our ad-hoc setting assumes a test-time partner is equally likely to follow any convention, we construct a $\gP^{(0)}_{\text{test}}$ population that is uniform over convention labels, using a hard-coded, single-fixed-action policy for each of the four conventions. We train best-response agents against the imbalanced population $\gP$ as well as convention-balanced populations $\gP'$ computed via matrix factorization and spectral clustering; for this experiment we additionally evaluate $k$-means clustering, applied directly to the rows of the cross-play matrix $\mX$, as a simpler alternative to the two.

\begin{figure}[h]
    \centering
    \includegraphics[width=0.8\textwidth]{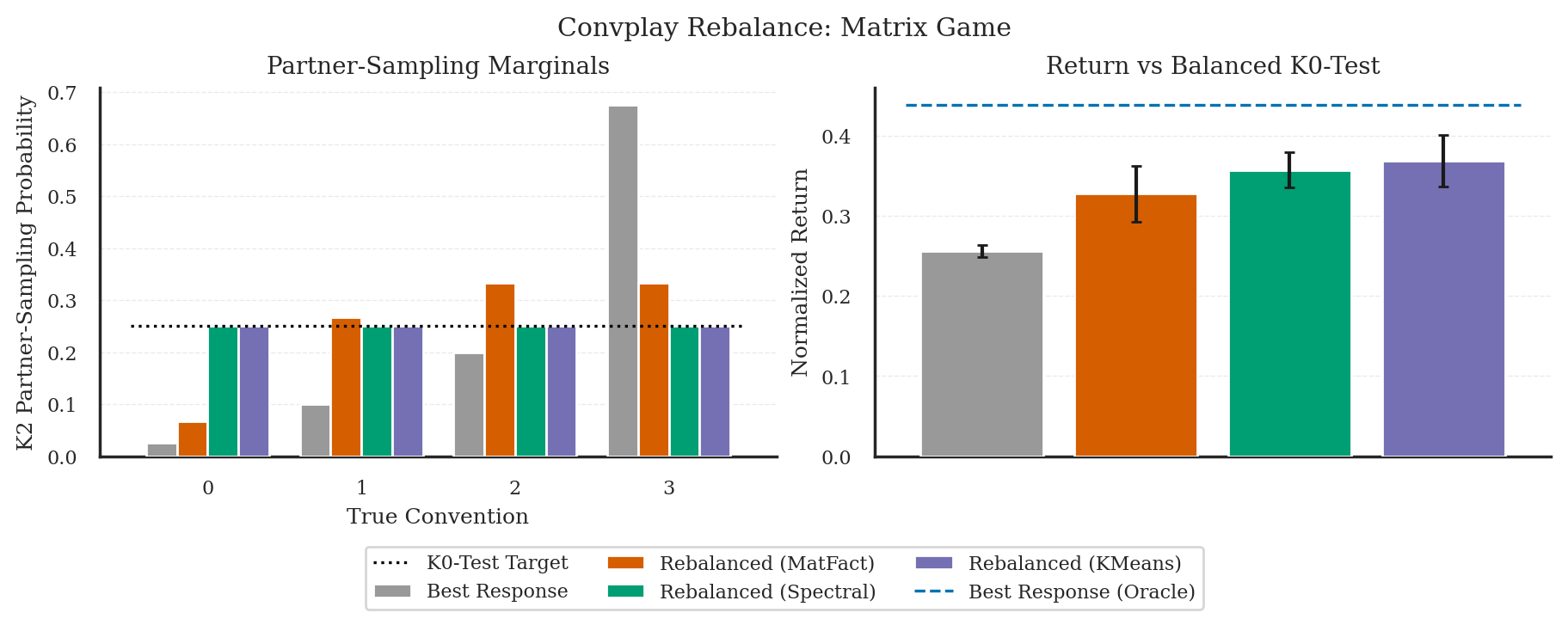}
    \caption{Left: $K_2$ partner-sampling marginal probability by true convention for the Matrix Game variant, comparing naive best-response sampling against matrix-factorization, spectral-clustering, and k-means rebalancing, relative to the uniform target (dotted line). Right: normalized episodic return for a best-response policy trained under each rebalancing scheme, evaluated against the uniform $\gP^{(0)}_{\text{test}}$ population, with an oracle upper bound (dashed line) given by the balanced test population's own self-crossplay.}
    \label{fig:rebalance_bars}
\end{figure}

\paragraph{Results} Figure~\ref{fig:rebalance_bars} reports both the partner-sampling marginals and the resulting returns for all best-response agents. Naive best-response sampling reproduces the population imbalance seen in Figure~\ref{fig:mx_commonE_unbalanced}, while spectral-clustering and k-means rebalancing both track the uniform target closely, indicating accurate discovery of the underlying convention system. Downstream performance mirrors this ordering: evaluated against the $\gP^{(0)}_{\text{test}}$ population, policies trained under spectral-clustering and k-means rebalancing achieve a significant 40\% improvement in return over the naive baseline.

\begin{figure}[h]
    \centering
    \includegraphics[width=0.7\textwidth]{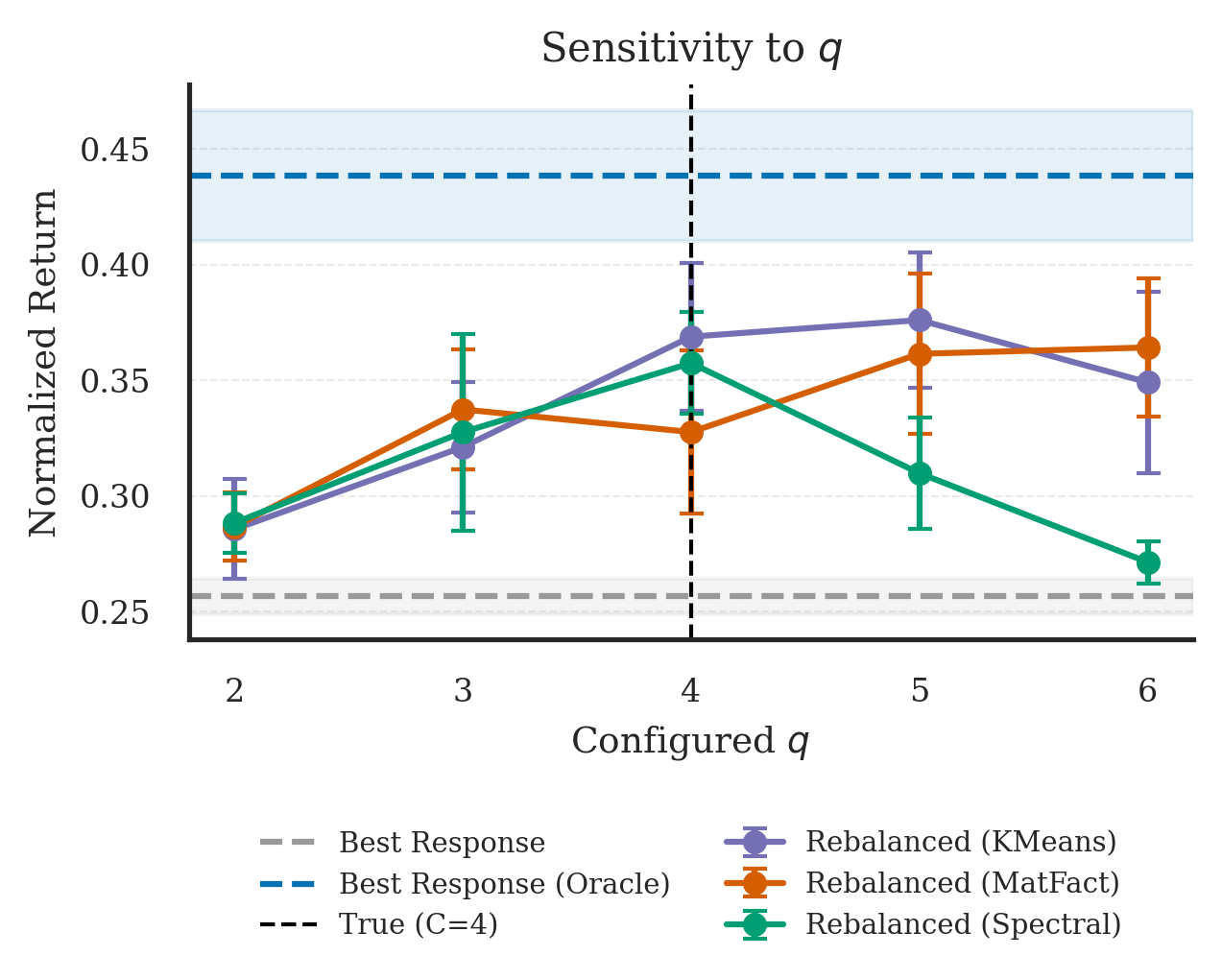}
    \caption{Normalized episodic return for a best-response policy trained under each rebalancing scheme (MatFact, Spectral, k-means), as a function of the specified number of conventions $q$, evaluated against the uniform $\gP^{(0)}_{\text{test}}$ population for the Matrix Game variant of Appendix~\ref{sec:rebalancing}. The dashed vertical line marks the ground-truth number of conventions. All methods improve over the unrebalanced baseline even when $q$ is mis-specified, with the largest gains at the correct value; MatFact remains comparatively resilient to overspecification of $q$.}
    \label{fig:rebalance_nc_ablation}
\end{figure}

We test the sensitivity of each partitioning method to its expected number of conventions, $q$. Figure~\ref{fig:rebalance_nc_ablation} reports normalized return under each method across a range of $q$ values spanning the ground-truth count. All three methods improve over the unrebalanced baseline across the full range tested, with performance maximized at the correct $q$. All methods degrade when $q$ is underspecified, but MatFact in particular remains resilient to overspecification. This sensitivity may vary across task domains, since the partitioning methods differ in how they handle non-transitive compatibility structure and noise in the cross-play matrix.

\subsection{Normalized Return}
\label{sec:normalized_return}

Reward values are normalized to each environment's known range, so results can be compared across experimental domains. Under the partial and full differentiation variants, many capability-limited agents cannot score the environment maximum, so the optimal score is not 100\%, but an average over each test agent's own achievable maximum. We compute this oracle empirically as the best response achievable against each test agent -- something we can compute directly since we know each agent's capabilities -- and plot the resulting average in Figure~\ref{fig:bar_reward_full} as a reference.

\KOneHeurRewardsTable

\subsection{Coordination Efficiency}
\label{sec:coord_efficiency}

To quantify how effectively an agent adapts to its partner's specific capability limitations, we introduce the \textit{Coordination Efficiency} $\eta(\pi, \pi')$ for a
policy $\pi$ against a partner $\pi'$, which we define as:
\begin{equation}
    \label{eqn:efficiency}
    \eta(\pi, \pi') = \frac{J(\pi, \pi')}{J^*(\pi')}
\end{equation}
where $J^*(\pi') = \max_{\hat{\pi} \in \Pi} J(\hat{\pi}, \pi')$ represents
the maximum possible return achievable with partner $\pi'$.  We approximate $J^*(\pi')$ based on the level $k$ of the partner strategy $\pi'$.  For a level $k=0$ partner, $J^*_{K_0}(\pi^{(0)}_i) = J_{SP}(\pi^{(0)}_i)$.  For a level $k=1$ partner, we define $J^*$ as the maximum joint return possible with any policy belonging to a convention within that partner's repertoire:
\begin{equation}
    J^*_{K_1}(\pi_i^{(1)}) = \max_{C \in \gK_i} \left[ \max_{\hat{\pi} \in C} J(\pi^{(1)}_i, \hat{\pi}) \right]
\end{equation}
This metric normalizes by each partner's achievable maximum return, providing a fair assessment of coordination efficiency even when conventions have substantially different expected returns.

\KZeroEfficiencyTable
\KOneEfficiencyTable
\KOneHeurEfficiencyTable

\subsection{Analysis of Probing in PMR}
\label{sec:pmr_probing}

\begin{figure}[h]
    \centering
    \includegraphics[width=\textwidth]{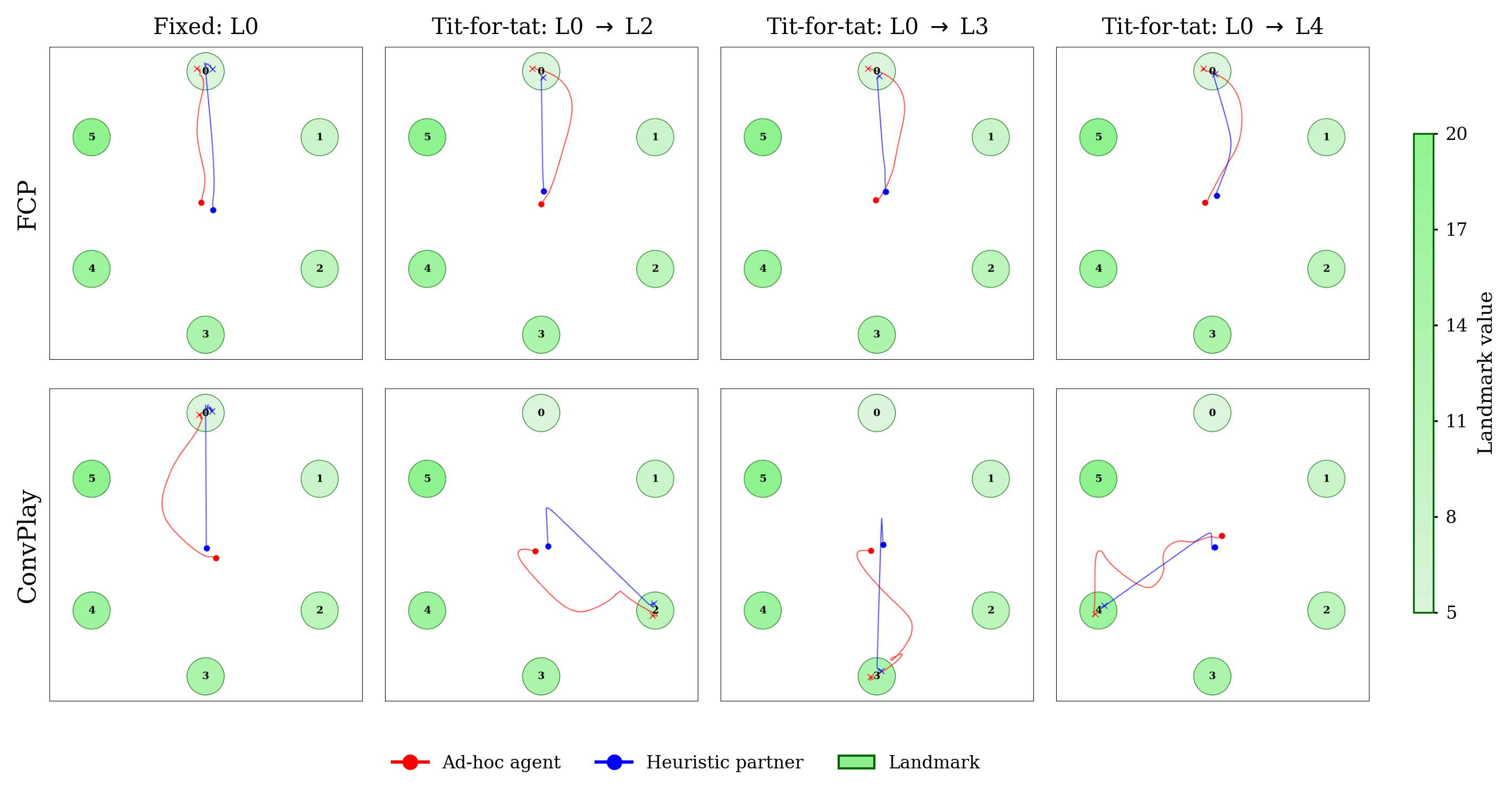}
    \caption{Trajectories in PMR for the ad-hoc agent (red) paired with a heuristic partner (blue), comparing FCP (top row) and \textit{ConventionPlay} (bottom row). Left column: a fixed $K_0$ partner committed to landmark 0. Remaining columns: tit-for-tat partners that default to landmark 0 before switching to landmark 2, 3, or 4 respectively, conditioned on the position of the partner agent. FCP agents infer their partner's initially preferred convention and coordinate accordingly. \textit{ConventionPlay}'s trajectories are visibly more winding, showing its ability to both lead a capability-limited adaptive partner or follow a fixed-convention one.}
    \label{fig:pmr_probe}
\end{figure}

Figure~\ref{fig:pmr_probe} shows \textit{ConventionPlay} against both a fixed-convention and adaptive partners in PMR. As in the matrix game, the ad-hoc agent ``probes'' with movement away from the partner's implied landmark and toward a higher-value one, in an attempt to induce the partner to adapt toward that convention.

\subsection{Effect of $K_1$ Population Size}
\label{sec:k1_size_effect}

\begin{figure}[h]
    \centering
    \includegraphics[width=0.9\columnwidth]{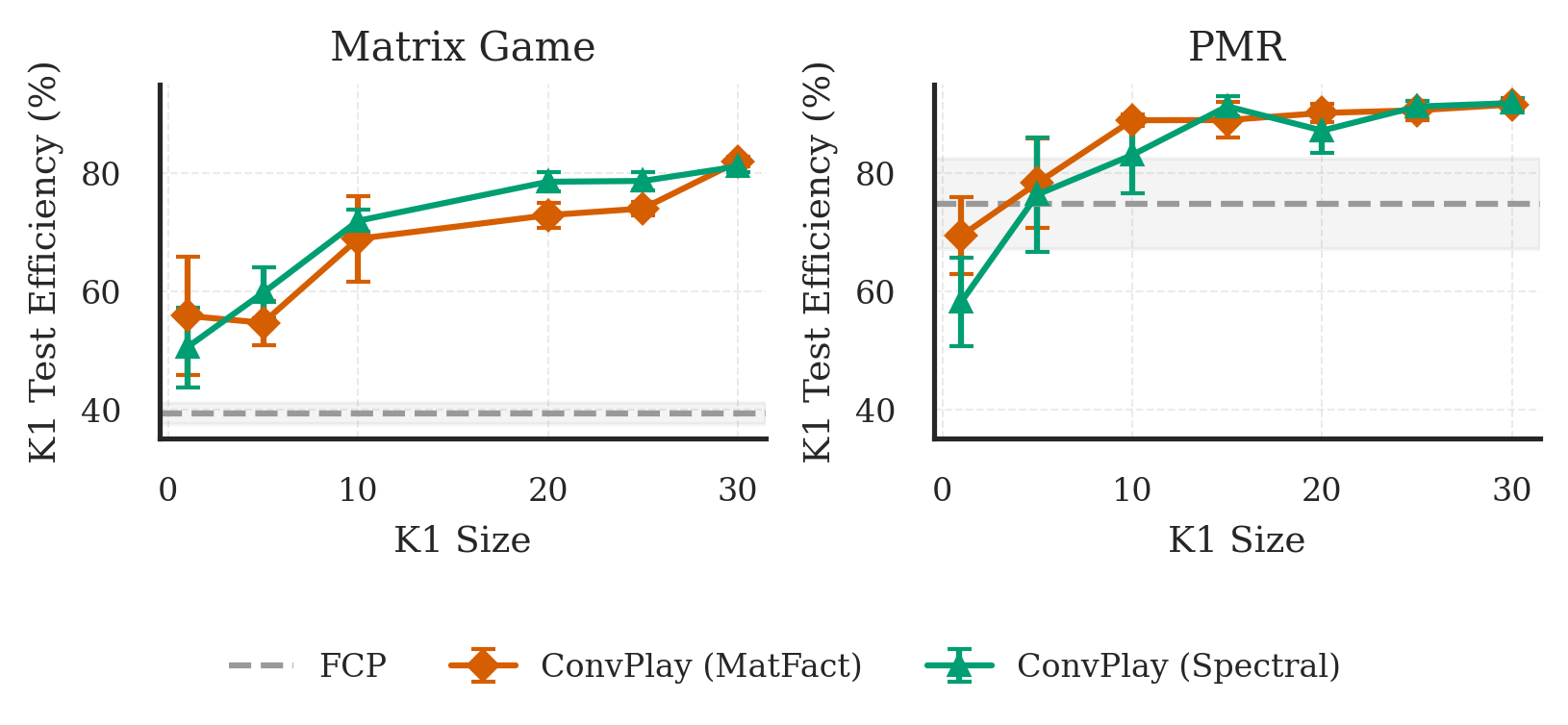}
    \caption{Effect of $K_1$ population size on coordination efficiency with $\gP^{(1)}_{\text{sampled}}$ partners. Coordination efficiency $\eta \in [0,1]$ is the fraction of a partner's achievable return obtained (defined in Appendix~\ref{sec:coord_efficiency}). \textit{ConventionPlay} exhibits substantial gains up to moderate diversity, followed by diminishing returns. FCP---invariant to $K_1$ size---is shown as a horizontal line for reference. Error bars show $\pm$1 sample standard deviation, where each seed's value is mean efficiency averaged over all test partners.
  }
\label{fig:k1_size_effect_v2}
\end{figure}

A critical design parameter is the size of the $K_1$ training population. With a low number of $K_1$ agents, \textit{ConventionPlay} should approximate naive best-response training. As population size increases, diversity of adaptive partners should enhance probing capability. Figure~\ref{fig:k1_size_effect_v2} reveals an inflection point in this relationship: performance gains are substantial up to a moderate $K_1$ size, after which additional diversity yields diminishing returns. The exact inflection point varies by environment and convention discovery method, suggesting complex interactions between convention decomposition approach and $K_1$ population size. A thorough analysis of these interactions remains an important direction for future work.

\end{document}

%% file: math_commands.tex
\usepackage{amsmath,amsfonts,bm}

\def\eqref#1{equation~\ref{#1}}

\def\1{\bm{1}}

\def\vw{{\bm{w}}}

\def\mA{{\bm{A}}}

\def\mE{{\bm{E}}}

\def\mW{{\bm{W}}}
\def\mX{{\bm{X}}}

\def\mZ{{\bm{Z}}}

\DeclareMathAlphabet{\mathsfit}{\encodingdefault}{\sfdefault}{m}{sl}
\SetMathAlphabet{\mathsfit}{bold}{\encodingdefault}{\sfdefault}{bx}{n}

\def\gA{{\mathcal{A}}}

\def\gC{{\mathcal{C}}}

\def\gH{{\mathcal{H}}}

\def\gK{{\mathcal{K}}}

\def\gM{{\mathcal{M}}}

\def\gP{{\mathcal{P}}}

\def\gS{{\mathcal{S}}}

\def\emA{{A}}

\def\emW{{W}}
\def\emX{{X}}

\def\emZ{{Z}}

\newcommand{\E}{\mathbb{E}}

\newcommand{\R}{\mathbb{R}}

\newcommand{\softmax}{\mathrm{softmax}}



%% file: iclr_data/convplay_tables.tex

\newcommand{\KZeroRewardsTable}{%
\begin{table}[h]
\centering
\caption{Mean normalised episodic return on $\gP^{(0)}_{\text{test}}$ partners. Each cell reports mean $\pm$ 1 s.d.\ (unbiased sample standard deviation, $n-1$ denominator) across $n=5$ seeds. Bold marks the row's best value.}
\label{tab:k0_rewards}
\resizebox{\columnwidth}{!}{%
\begin{tabular}{llccccccccc}
\toprule
Environment & Var & FCP & SyKLRBR (L5) & SyKLRBR (L2) & CoMeDi (BC) & CoMeDi (BR) & LIPO & MEP & ConvPlay (matfact) & ConvPlay (spectral) \\
\midrule
Matrix Game & Non-diff & 95.16 $\pm$ 0.30 & 95.47 $\pm$ 0.17 & 95.55 $\pm$ 0.14 & 67.64 $\pm$ 0.21 & 68.86 $\pm$ 4.86 & \textbf{95.58 $\pm$ 0.09} & 69.44 $\pm$ 1.31 & 94.64 $\pm$ 0.15 & 94.42 $\pm$ 0.37 \\
 & Partial & 41.76 $\pm$ 0.14 & 41.76 $\pm$ 0.20 & \textbf{41.82 $\pm$ 0.20} & 31.45 $\pm$ 0.11 & 30.57 $\pm$ 2.75 & 38.70 $\pm$ 0.10 & 34.18 $\pm$ 3.36 & 40.69 $\pm$ 1.31 & 41.21 $\pm$ 0.16 \\
 & Full & \textbf{37.44 $\pm$ 0.06} & 37.23 $\pm$ 0.76 & 37.42 $\pm$ 0.52 & 29.00 $\pm$ 0.09 & 29.95 $\pm$ 0.67 & 33.90 $\pm$ 0.74 & 30.25 $\pm$ 0.83 & 36.35 $\pm$ 0.44 & 36.56 $\pm$ 0.57 \\
\midrule
PMR & Non-diff & 89.42 $\pm$ 7.94 & \textbf{97.54 $\pm$ 0.24} & 97.46 $\pm$ 0.14 & 66.48 $\pm$ 3.18 & 80.17 $\pm$ 11.29 & 90.64 $\pm$ 6.80 & 52.11 $\pm$ 7.26 & 78.21 $\pm$ 7.14 & 91.07 $\pm$ 9.01 \\
 & Partial & 60.33 $\pm$ 6.97 & \textbf{66.00 $\pm$ 0.15} & 65.98 $\pm$ 0.19 & 35.49 $\pm$ 5.25 & 53.16 $\pm$ 6.44 & 43.60 $\pm$ 0.27 & 53.96 $\pm$ 10.76 & 64.85 $\pm$ 2.12 & 65.80 $\pm$ 0.26 \\
 & Full & 56.49 $\pm$ 6.22 & 54.27 $\pm$ 6.05 & 32.68 $\pm$ 7.56 & 16.56 $\pm$ 7.40 & 37.81 $\pm$ 10.94 & 47.17 $\pm$ 7.54 & 38.17 $\pm$ 2.51 & 57.92 $\pm$ 3.62 & \textbf{60.58 $\pm$ 0.11} \\
\midrule
Task Relay & Non-diff & \textbf{90.00 $\pm$ 0.87} & 87.56 $\pm$ 3.60 & 87.06 $\pm$ 4.79 & 62.93 $\pm$ 4.57 & 38.21 $\pm$ 9.79 & 74.11 $\pm$ 5.43 & 78.92 $\pm$ 4.52 & 79.36 $\pm$ 3.55 & 83.12 $\pm$ 6.32 \\
 & Partial & 78.17 $\pm$ 1.20 & 80.56 $\pm$ 2.57 & \textbf{81.31 $\pm$ 0.81} & 46.77 $\pm$ 9.67 & 61.41 $\pm$ 7.56 & 57.39 $\pm$ 1.30 & 68.14 $\pm$ 3.52 & 70.83 $\pm$ 4.04 & 74.28 $\pm$ 6.15 \\
 & Full & 38.88 $\pm$ 9.05 & 44.26 $\pm$ 4.43 & 42.68 $\pm$ 5.95 & 16.84 $\pm$ 1.35 & 19.86 $\pm$ 6.65 & 18.88 $\pm$ 4.90 & 3.10 $\pm$ 1.55 & 31.13 $\pm$ 12.45 & \textbf{44.70 $\pm$ 4.44} \\
\bottomrule
\end{tabular}%
}
\end{table}
}

\newcommand{\KOneRewardsTable}{%
\begin{table}[h]
\centering
\caption{Mean normalised episodic return on $\gP^{(1)}_{\text{sampled}}$ partners. Each cell reports mean $\pm$ 1 s.d.\ (unbiased sample standard deviation, $n-1$ denominator) across $n=5$ seeds. Bold marks the row's best value.}
\label{tab:k1_rewards}
\resizebox{\columnwidth}{!}{%
\begin{tabular}{llccccccccc}
\toprule
Environment & Var & FCP & SyKLRBR (L5) & SyKLRBR (L2) & CoMeDi (BC) & CoMeDi (BR) & LIPO & MEP & ConvPlay (matfact) & ConvPlay (spectral) \\
\midrule
Matrix Game & Non-diff & 65.93 $\pm$ 2.69 & 63.78 $\pm$ 2.06 & 63.64 $\pm$ 2.56 & 66.48 $\pm$ 1.59 & 70.66 $\pm$ 3.31 & 60.41 $\pm$ 5.12 & 71.98 $\pm$ 4.02 & \textbf{86.19 $\pm$ 1.76} & 85.52 $\pm$ 1.91 \\
 & Partial & 26.03 $\pm$ 3.90 & 25.65 $\pm$ 2.41 & 25.56 $\pm$ 2.50 & 47.05 $\pm$ 0.21 & 38.23 $\pm$ 1.58 & 30.09 $\pm$ 0.91 & 39.86 $\pm$ 2.31 & 46.50 $\pm$ 0.93 & \textbf{50.47 $\pm$ 1.52} \\
 & Full & 24.03 $\pm$ 1.31 & 25.74 $\pm$ 4.37 & 23.97 $\pm$ 4.24 & 48.12 $\pm$ 0.33 & 33.33 $\pm$ 1.22 & 33.47 $\pm$ 2.68 & 46.27 $\pm$ 0.89 & 51.52 $\pm$ 1.37 & \textbf{52.63 $\pm$ 0.41} \\
\midrule
PMR & Non-diff & 89.33 $\pm$ 5.75 & 94.39 $\pm$ 0.37 & \textbf{94.43 $\pm$ 0.52} & 63.70 $\pm$ 3.95 & 80.00 $\pm$ 8.47 & 90.10 $\pm$ 5.53 & 62.91 $\pm$ 8.20 & 86.30 $\pm$ 4.50 & 93.46 $\pm$ 2.14 \\
 & Partial & 58.44 $\pm$ 5.35 & 62.25 $\pm$ 1.10 & 62.93 $\pm$ 1.93 & 40.57 $\pm$ 3.44 & 57.89 $\pm$ 7.25 & 47.82 $\pm$ 1.19 & 56.30 $\pm$ 8.30 & \textbf{74.82 $\pm$ 1.16} & 72.58 $\pm$ 1.24 \\
 & Full & 53.73 $\pm$ 5.81 & 53.08 $\pm$ 4.95 & 36.07 $\pm$ 3.68 & 20.99 $\pm$ 10.27 & 39.29 $\pm$ 9.71 & 47.62 $\pm$ 5.66 & 43.74 $\pm$ 2.96 & 65.54 $\pm$ 1.41 & \textbf{66.11 $\pm$ 1.28} \\
\midrule
Task Relay & Non-diff & 82.72 $\pm$ 2.43 & 8.56 $\pm$ 8.64 & 5.90 $\pm$ 6.51 & 53.84 $\pm$ 9.61 & 2.40 $\pm$ 4.65 & 7.74 $\pm$ 9.18 & 18.17 $\pm$ 20.86 & 75.09 $\pm$ 21.41 & \textbf{83.22 $\pm$ 4.52} \\
 & Partial & 43.40 $\pm$ 36.42 & 2.67 $\pm$ 0.90 & 2.88 $\pm$ 1.08 & 41.34 $\pm$ 5.81 & 53.76 $\pm$ 27.75 & 1.62 $\pm$ 1.41 & 22.83 $\pm$ 14.01 & 62.81 $\pm$ 8.21 & \textbf{75.12 $\pm$ 11.31} \\
 & Full & 14.45 $\pm$ 23.57 & 3.30 $\pm$ 3.66 & 2.29 $\pm$ 1.94 & 8.93 $\pm$ 1.14 & 4.43 $\pm$ 3.53 & 0.56 $\pm$ 0.43 & 1.18 $\pm$ 0.42 & 32.38 $\pm$ 20.70 & \textbf{53.66 $\pm$ 8.03} \\
\bottomrule
\end{tabular}%
}
\end{table}
}

\newcommand{\KZeroEfficiencyTable}{%
\begin{table}[h]
\centering
\caption{Coordination efficiency $\eta$ (\%) on $\gP^{(0)}_{\text{test}}$ partners. Each cell reports mean $\pm$ 1 s.d.\ (unbiased sample standard deviation, $n-1$ denominator) across $n=5$ seeds. Bold marks the row's best value.}
\label{tab:k0_efficiency}
\resizebox{\columnwidth}{!}{%
\begin{tabular}{llccccccccc}
\toprule
Environment & Var & FCP & SyKLRBR (L5) & SyKLRBR (L2) & CoMeDi (BC) & CoMeDi (BR) & LIPO & MEP & ConvPlay (matfact) & ConvPlay (spectral) \\
\midrule
Matrix Game & Non-diff & 95.16 $\pm$ 0.30 & 95.47 $\pm$ 0.17 & 95.55 $\pm$ 0.14 & 67.64 $\pm$ 0.21 & 68.86 $\pm$ 4.86 & \textbf{95.58 $\pm$ 0.09} & 69.44 $\pm$ 1.31 & 94.64 $\pm$ 0.15 & 94.42 $\pm$ 0.37 \\
 & Partial & 95.35 $\pm$ 0.55 & 95.25 $\pm$ 0.35 & \textbf{95.41 $\pm$ 0.40} & 55.77 $\pm$ 0.24 & 53.83 $\pm$ 6.64 & 83.15 $\pm$ 0.30 & 68.33 $\pm$ 10.25 & 90.19 $\pm$ 5.04 & 92.12 $\pm$ 0.37 \\
 & Full & \textbf{95.26 $\pm$ 0.26} & 92.53 $\pm$ 5.88 & 93.07 $\pm$ 5.35 & 44.80 $\pm$ 0.16 & 48.01 $\pm$ 1.97 & 64.11 $\pm$ 4.55 & 56.18 $\pm$ 0.47 & 87.36 $\pm$ 4.78 & 86.02 $\pm$ 5.34 \\
\midrule
PMR & Non-diff & 90.05 $\pm$ 7.96 & \textbf{98.20 $\pm$ 0.24} & 98.11 $\pm$ 0.14 & 66.95 $\pm$ 3.24 & 80.55 $\pm$ 11.43 & 91.27 $\pm$ 6.79 & 52.50 $\pm$ 7.26 & 78.71 $\pm$ 7.25 & 91.70 $\pm$ 9.04 \\
 & Partial & 90.14 $\pm$ 8.64 & \textbf{98.20 $\pm$ 0.13} & 98.16 $\pm$ 0.19 & 51.02 $\pm$ 7.01 & 73.20 $\pm$ 8.43 & 63.66 $\pm$ 0.32 & 83.68 $\pm$ 13.72 & 94.31 $\pm$ 7.10 & 97.53 $\pm$ 0.24 \\
 & Full & 87.40 $\pm$ 13.70 & 88.43 $\pm$ 8.62 & 78.20 $\pm$ 10.79 & 30.13 $\pm$ 9.03 & 63.62 $\pm$ 14.91 & 70.07 $\pm$ 8.67 & 55.67 $\pm$ 9.16 & 90.22 $\pm$ 8.88 & \textbf{96.69 $\pm$ 0.21} \\
\midrule
Task Relay & Non-diff & \textbf{91.49 $\pm$ 0.90} & 89.00 $\pm$ 3.68 & 88.49 $\pm$ 4.86 & 64.06 $\pm$ 4.67 & 38.93 $\pm$ 9.97 & 75.44 $\pm$ 5.45 & 80.13 $\pm$ 4.60 & 80.65 $\pm$ 3.57 & 84.48 $\pm$ 6.45 \\
 & Partial & \textbf{95.65 $\pm$ 7.99} & 92.18 $\pm$ 3.17 & 94.28 $\pm$ 2.23 & 55.15 $\pm$ 11.91 & 72.27 $\pm$ 9.88 & 65.41 $\pm$ 1.47 & 79.46 $\pm$ 4.58 & 81.92 $\pm$ 4.61 & 89.24 $\pm$ 7.00 \\
 & Full & 53.37 $\pm$ 13.16 & 60.69 $\pm$ 5.83 & 59.61 $\pm$ 7.35 & 27.14 $\pm$ 2.35 & 32.49 $\pm$ 9.79 & 29.78 $\pm$ 6.19 & 8.05 $\pm$ 1.94 & 42.80 $\pm$ 18.10 & \textbf{62.90 $\pm$ 4.85} \\
\bottomrule
\end{tabular}%
}
\end{table}
}

\newcommand{\KOneEfficiencyTable}{%
\begin{table}[h]
\centering
\caption{Coordination efficiency $\eta$ (\%) on $\gP^{(1)}_{\text{sampled}}$ partners. Each cell reports mean $\pm$ 1 s.d.\ (unbiased sample standard deviation, $n-1$ denominator) across $n=5$ seeds. Bold marks the row's best value.}
\label{tab:k1_efficiency}
\resizebox{\columnwidth}{!}{%
\begin{tabular}{llccccccccc}
\toprule
Environment & Var & FCP & SyKLRBR (L5) & SyKLRBR (L2) & CoMeDi (BC) & CoMeDi (BR) & LIPO & MEP & ConvPlay (matfact) & ConvPlay (spectral) \\
\midrule
Matrix Game & Non-diff & 66.88 $\pm$ 2.76 & 64.68 $\pm$ 2.10 & 64.54 $\pm$ 2.61 & 67.45 $\pm$ 1.62 & 71.73 $\pm$ 3.36 & 61.28 $\pm$ 5.21 & 73.10 $\pm$ 4.09 & \textbf{87.47 $\pm$ 1.79} & 86.81 $\pm$ 1.93 \\
 & Partial & 46.69 $\pm$ 6.71 & 46.24 $\pm$ 3.65 & 46.04 $\pm$ 4.17 & 67.40 $\pm$ 0.44 & 59.71 $\pm$ 3.32 & 53.29 $\pm$ 1.11 & 63.24 $\pm$ 5.07 & 76.38 $\pm$ 1.84 & \textbf{84.23 $\pm$ 1.82} \\
 & Full & 39.32 $\pm$ 1.84 & 42.34 $\pm$ 6.18 & 39.32 $\pm$ 6.25 & 59.74 $\pm$ 0.52 & 50.83 $\pm$ 1.96 & 51.35 $\pm$ 4.97 & 64.18 $\pm$ 2.11 & 73.99 $\pm$ 1.15 & \textbf{81.11 $\pm$ 0.92} \\
\midrule
PMR & Non-diff & 93.71 $\pm$ 6.04 & 99.02 $\pm$ 0.39 & \textbf{99.07 $\pm$ 0.54} & 66.82 $\pm$ 4.12 & 83.90 $\pm$ 8.91 & 94.51 $\pm$ 5.83 & 65.86 $\pm$ 8.60 & 90.53 $\pm$ 4.72 & 98.03 $\pm$ 2.26 \\
 & Partial & 73.24 $\pm$ 5.58 & 77.58 $\pm$ 1.22 & 78.43 $\pm$ 2.32 & 47.97 $\pm$ 5.34 & 71.37 $\pm$ 7.19 & 60.96 $\pm$ 1.51 & 70.84 $\pm$ 9.72 & \textbf{92.03 $\pm$ 1.65} & 90.17 $\pm$ 1.54 \\
 & Full & 74.78 $\pm$ 7.65 & 73.55 $\pm$ 6.69 & 52.59 $\pm$ 5.85 & 31.71 $\pm$ 8.82 & 55.86 $\pm$ 11.29 & 67.04 $\pm$ 7.07 & 59.92 $\pm$ 3.88 & 88.94 $\pm$ 3.00 & \textbf{91.27 $\pm$ 1.73} \\
\midrule
Task Relay & Non-diff & 85.43 $\pm$ 2.63 & 8.78 $\pm$ 8.82 & 6.08 $\pm$ 6.61 & 55.49 $\pm$ 10.09 & 2.47 $\pm$ 4.76 & 7.91 $\pm$ 9.35 & 18.91 $\pm$ 21.60 & 77.75 $\pm$ 22.09 & \textbf{86.07 $\pm$ 4.80} \\
 & Partial & 47.00 $\pm$ 39.51 & 2.83 $\pm$ 0.99 & 3.23 $\pm$ 1.07 & 43.84 $\pm$ 6.61 & 58.08 $\pm$ 30.14 & 1.63 $\pm$ 1.41 & 24.55 $\pm$ 15.87 & 68.19 $\pm$ 8.59 & \textbf{81.16 $\pm$ 11.96} \\
 & Full & 18.17 $\pm$ 29.50 & 4.31 $\pm$ 4.32 & 2.83 $\pm$ 2.35 & 10.78 $\pm$ 1.76 & 5.10 $\pm$ 3.91 & 0.76 $\pm$ 0.57 & 1.55 $\pm$ 0.55 & 40.14 $\pm$ 26.49 & \textbf{67.35 $\pm$ 10.91} \\
\bottomrule
\end{tabular}%
}
\end{table}
}

\newcommand{\KOneHeurRewardsTable}{%
\begin{table}[h]
\centering
\caption{Mean normalised episodic return on $\gP^{(1)}_{\text{heur}}$ partners. Each cell reports mean $\pm$ 1 s.d.\ (unbiased sample standard deviation, $n-1$ denominator) across $n=5$ seeds. Bold marks the row's best value.}
\label{tab:k1_heur_rewards}
\resizebox{\columnwidth}{!}{%
\begin{tabular}{llccccccccc}
\toprule
Environment & Var & FCP & SyKLRBR (L5) & SyKLRBR (L2) & CoMeDi (BC) & CoMeDi (BR) & LIPO & MEP & ConvPlay (matfact) & ConvPlay (spectral) \\
\midrule
Matrix Game & Non-diff & 93.04 $\pm$ 0.83 & 93.47 $\pm$ 0.43 & \textbf{94.06 $\pm$ 0.50} & 44.25 $\pm$ 1.88 & 87.25 $\pm$ 2.39 & 94.02 $\pm$ 0.38 & 81.26 $\pm$ 18.89 & 93.62 $\pm$ 0.49 & 93.01 $\pm$ 0.68 \\
 & Partial & 27.07 $\pm$ 0.98 & 27.77 $\pm$ 1.07 & 27.75 $\pm$ 0.92 & 9.11 $\pm$ 0.32 & 13.29 $\pm$ 1.66 & 27.60 $\pm$ 0.54 & 22.85 $\pm$ 5.02 & 28.75 $\pm$ 4.37 & \textbf{29.54 $\pm$ 0.60} \\
 & Full & 14.87 $\pm$ 0.58 & 15.38 $\pm$ 2.55 & 15.91 $\pm$ 0.96 & 8.69 $\pm$ 0.17 & 12.97 $\pm$ 1.12 & 14.36 $\pm$ 2.18 & 14.25 $\pm$ 1.00 & \textbf{17.60 $\pm$ 2.15} & 17.24 $\pm$ 1.46 \\
\midrule
PMR & Non-diff & 62.91 $\pm$ 6.71 & 68.17 $\pm$ 1.21 & 68.30 $\pm$ 0.93 & 51.17 $\pm$ 3.28 & 63.83 $\pm$ 7.42 & \textbf{68.52 $\pm$ 1.44} & 47.63 $\pm$ 2.14 & 60.78 $\pm$ 4.33 & 64.94 $\pm$ 7.64 \\
 & Partial & 32.49 $\pm$ 3.39 & 32.18 $\pm$ 1.03 & 32.73 $\pm$ 1.06 & 21.83 $\pm$ 1.91 & 28.78 $\pm$ 3.08 & 26.10 $\pm$ 1.30 & 30.84 $\pm$ 4.90 & 39.18 $\pm$ 2.49 & \textbf{39.26 $\pm$ 1.33} \\
 & Full & 27.88 $\pm$ 4.42 & 27.52 $\pm$ 2.34 & 27.20 $\pm$ 2.77 & 7.96 $\pm$ 7.46 & 23.64 $\pm$ 2.72 & 25.84 $\pm$ 1.61 & 19.74 $\pm$ 3.37 & 34.42 $\pm$ 1.48 & \textbf{36.64 $\pm$ 1.50} \\
\midrule
Task Relay & Non-diff & \textbf{76.91 $\pm$ 9.41} & 3.54 $\pm$ 5.71 & 1.70 $\pm$ 2.68 & 43.17 $\pm$ 13.17 & 12.10 $\pm$ 11.38 & 5.83 $\pm$ 7.65 & 13.51 $\pm$ 9.66 & 68.91 $\pm$ 26.33 & 70.79 $\pm$ 6.58 \\
 & Partial & 16.21 $\pm$ 19.55 & 0.00 $\pm$ 0.00 & 1.13 $\pm$ 1.28 & 40.23 $\pm$ 9.31 & 39.47 $\pm$ 21.22 & 0.01 $\pm$ 0.02 & 23.28 $\pm$ 13.24 & 45.46 $\pm$ 31.67 & \textbf{68.73 $\pm$ 1.91} \\
 & Full & 24.80 $\pm$ 20.81 & 10.06 $\pm$ 5.76 & 13.62 $\pm$ 4.50 & 23.41 $\pm$ 1.24 & 13.86 $\pm$ 5.19 & 0.44 $\pm$ 0.79 & 3.37 $\pm$ 2.91 & 50.93 $\pm$ 21.21 & \textbf{58.60 $\pm$ 12.17} \\
\bottomrule
\end{tabular}%
}
\end{table}
}

\newcommand{\KOneHeurEfficiencyTable}{%
\begin{table}[h]
\centering
\caption{Coordination efficiency $\eta$ (\%) on $\gP^{(1)}_{\text{heur}}$ partners. Each cell reports mean $\pm$ 1 s.d.\ (unbiased sample standard deviation, $n-1$ denominator) across $n=5$ seeds. Bold marks the row's best value.}
\label{tab:k1_heur_efficiency}
\resizebox{\columnwidth}{!}{%
\begin{tabular}{llccccccccc}
\toprule
Environment & Var & FCP & SyKLRBR (L5) & SyKLRBR (L2) & CoMeDi (BC) & CoMeDi (BR) & LIPO & MEP & ConvPlay (matfact) & ConvPlay (spectral) \\
\midrule
Matrix Game & Non-diff & 93.77 $\pm$ 0.83 & 94.22 $\pm$ 0.43 & \textbf{94.80 $\pm$ 0.50} & 44.62 $\pm$ 1.89 & 87.93 $\pm$ 2.41 & 94.76 $\pm$ 0.39 & 81.90 $\pm$ 19.03 & 94.36 $\pm$ 0.49 & 93.74 $\pm$ 0.68 \\
 & Partial & 73.68 $\pm$ 1.64 & 74.50 $\pm$ 1.32 & 74.66 $\pm$ 1.25 & 10.26 $\pm$ 0.67 & 20.37 $\pm$ 4.23 & 74.51 $\pm$ 0.63 & 56.05 $\pm$ 15.59 & 68.33 $\pm$ 13.38 & \textbf{75.19 $\pm$ 1.00} \\
 & Full & 49.06 $\pm$ 1.02 & 46.20 $\pm$ 9.34 & 48.91 $\pm$ 4.48 & 9.38 $\pm$ 0.31 & 26.12 $\pm$ 2.95 & 27.99 $\pm$ 3.60 & 32.67 $\pm$ 3.80 & 42.12 $\pm$ 8.41 & \textbf{51.01 $\pm$ 3.26} \\
\midrule
PMR & Non-diff & 86.32 $\pm$ 9.15 & 93.47 $\pm$ 1.66 & 93.68 $\pm$ 1.30 & 70.08 $\pm$ 4.49 & 87.45 $\pm$ 10.39 & \textbf{93.94 $\pm$ 1.96} & 64.77 $\pm$ 2.91 & 83.32 $\pm$ 5.95 & 88.98 $\pm$ 10.65 \\
 & Partial & 65.43 $\pm$ 7.36 & 65.73 $\pm$ 1.84 & 66.81 $\pm$ 1.76 & 36.71 $\pm$ 4.63 & 55.79 $\pm$ 6.80 & 52.83 $\pm$ 2.26 & 63.85 $\pm$ 8.86 & 75.41 $\pm$ 6.62 & \textbf{75.83 $\pm$ 2.72} \\
 & Full & 62.86 $\pm$ 9.40 & 62.71 $\pm$ 4.36 & 61.84 $\pm$ 5.23 & 19.93 $\pm$ 18.25 & 53.51 $\pm$ 6.56 & 57.06 $\pm$ 3.58 & 44.10 $\pm$ 7.84 & 73.11 $\pm$ 3.92 & \textbf{77.68 $\pm$ 3.22} \\
\midrule
Task Relay & Non-diff & \textbf{81.55 $\pm$ 10.06} & 3.73 $\pm$ 6.03 & 1.79 $\pm$ 2.82 & 45.73 $\pm$ 13.94 & 12.75 $\pm$ 11.99 & 6.23 $\pm$ 8.15 & 14.33 $\pm$ 10.32 & 73.03 $\pm$ 27.95 & 75.18 $\pm$ 6.93 \\
 & Partial & 17.89 $\pm$ 21.18 & 0.00 $\pm$ 0.00 & 1.26 $\pm$ 1.47 & 43.66 $\pm$ 9.86 & 42.64 $\pm$ 23.63 & 0.01 $\pm$ 0.02 & 25.38 $\pm$ 15.08 & 50.41 $\pm$ 35.06 & \textbf{76.05 $\pm$ 2.01} \\
 & Full & 28.44 $\pm$ 24.40 & 11.19 $\pm$ 6.64 & 15.47 $\pm$ 5.14 & 28.50 $\pm$ 1.55 & 16.69 $\pm$ 6.46 & 0.48 $\pm$ 0.82 & 3.91 $\pm$ 3.33 & 59.92 $\pm$ 25.07 & \textbf{68.05 $\pm$ 14.54} \\
\bottomrule
\end{tabular}%
}
\end{table}
}


%% file: iclr_data/rebalance_tables.tex

\newcommand{\RebalanceNormalizedRewardTable}{%
\begin{table}[h]
\centering
\caption{Mean normalized ($[0, 1]$) episodic return on $K_0$ test partners, floor/ceiling set by the balanced $K_0$-test population's own self-crossplay. Each cell reports mean $\pm$ 1 s.d.\ across $n=5$ seeds. Bold marks the row's best value.}
\label{tab:rebalance_normalized_reward}
\resizebox{\columnwidth}{!}{%
\begin{tabular}{lccccc}
\toprule
Environment & Best Response & Rebalanced (MatFact) & Rebalanced (Spectral) & Rebalanced (KMeans) & Oracle \\
\midrule
Matrix Game & 0.26 $\pm$ 0.01 & 0.33 $\pm$ 0.04 & 0.36 $\pm$ 0.02 & \textbf{0.37 $\pm$ 0.03} & 0.44 \\
\bottomrule
\end{tabular}%
}
\end{table}
}


%% file: iclr_data/hyperparams_tables.tex

\newcommand{\SelfplayHyperparamsTable}{%
\begin{table}[h]
\centering
\caption{Self-play hyperparameters by environment family.}
\label{tab:selfplay-hyperparams}
\adjustbox{max width=\columnwidth}{%
\small
\begin{tabular}{lccc}
\toprule
Hyperparameter & Matrix Game & PMR & Task Relay \\
\midrule
\multicolumn{4}{l}{\textit{Training Loop}} \\
Total Env Frames & 12,500 & $1 \times 10^{7}$ & $1.5 \times 10^{7}$ \\
Training Iterations & 25 & 50 & 100 \\
Frames per Iteration & 500 & $2 \times 10^{5}$ & $1.5 \times 10^{5}$ \\
Minibatch Size & 500 & 4,000 & 1,500 \\
Minibatches per Iteration & 10 & 50 & 100 \\
Epochs per Iteration & 1.00 & 1.00 & 1.00 \\
Buffer Size & 5,000 & $2 \times 10^{5}$ & $1.5 \times 10^{5}$ \\
Sequence Length & -- & -- & -- \\
Max Grad Norm & 10 & 2 & 10 \\
\midrule
\multicolumn{4}{l}{\textit{Learner}} \\
Algorithm & IQL & PPO & PPO \\
Policy Network & MLP [32, 32] & MLP [256, 256] & MLP [256, 256] \\
Critic Network & -- & MLP [256, 256] & MLP [256, 256] \\
Learning Rate (LR) & 0.001 & 0.0009 & 0.0009 \\
Adam Epsilon & 0.001 & 1e-05 & 1e-05 \\
LR Annealed & Yes & Yes & Yes \\
Discount Factor ($\gamma$) & 0.99 & 0.99 & 0.99 \\
GAE $\lambda$ & -- & 0.95 & 0.95 \\
Clip Epsilon & -- & 0.2 & 0.2 \\
Entropy Coefficient & -- & 0.01 & 0.01 \\
Critic Loss Coefficient & -- & 0.5 & 0.5 \\
Epsilon-Greedy Init & 0.8 & -- & -- \\
Epsilon-Greedy End & 0.05 & -- & -- \\
Exploration Fraction & 0.5 & -- & -- \\
Target Update $\tau$ & 0.05 & -- & -- \\
Delayed Target Network & Yes & -- & -- \\
TD($\lambda$) & 0 & -- & -- \\
\bottomrule
\end{tabular}%
}
\end{table}
}

\newcommand{\AdaptiveplayHyperparamsTable}{%
\begin{table}[h]
\centering
\caption{Best-response hyperparameters by environment family.}
\label{tab:adaptiveplay-hyperparams}
\adjustbox{max width=\columnwidth}{%
\small
\begin{tabular}{lccc}
\toprule
Hyperparameter & Matrix Game & PMR & Task Relay \\
\midrule
\multicolumn{4}{l}{\textit{Training Loop}} \\
Total Env Frames & $1.6 \times 10^{6}$ & $4 \times 10^{6}$ & $7.5 \times 10^{6}$ \\
Training Iterations & 20 & 5 & 50 \\
Frames per Iteration & 80,000 & $8 \times 10^{5}$ & $1.5 \times 10^{5}$ \\
Minibatch Size & 5,000 & 8,000 & 2,000 \\
Minibatches per Iteration & 20 & 100 & 75 \\
Epochs per Iteration & 1.25 & 1.00 & 1.00 \\
Buffer Size & 80,000 & $8 \times 10^{5}$ & $1.5 \times 10^{5}$ \\
Sequence Length & 5 & 10 & 10 \\
Max Grad Norm & 10 & 2 & 10 \\
\midrule
\multicolumn{4}{l}{\textit{Learner}} \\
Algorithm & PPO & PPO & PPO \\
Policy Network & \shortstack{MLP [64, 64] $\to$ 64 \\ $\to$ LSTM(128) $\to$ MLP [64, 64]} & \shortstack{MLP [256, 256] $\to$ 128 \\ $\to$ LSTM(128) $\to$ MLP [256, 256]} & \shortstack{MLP [64, 64] $\to$ 64 \\ $\to$ LSTM(128) $\to$ MLP [64, 64]} \\
Critic Network & MLP [256, 256] & MLP [256, 256] & MLP [256, 256] \\
Learning Rate (LR) & 0.005 & 0.001 & 0.005 \\
Adam Epsilon & 1e-05 & 1e-05 & 1e-05 \\
LR Annealed & Yes & Yes & Yes \\
Discount Factor ($\gamma$) & 0.99 & 0.99 & 0.99 \\
GAE $\lambda$ & 0.95 & 0.95 & 0.95 \\
Clip Epsilon & 0.2 & 0.3 & 0.2 \\
Entropy Coefficient & 0.1 & 0.25 & 0.1 \\
Critic Loss Coefficient & 0.5 & 0.5 & 0.5 \\
\bottomrule
\end{tabular}%
}
\end{table}
}


%% file: all_items.bib
@inproceedings{10.5555/2980539.2980649,
author = {Ng, Andrew Y. and Jordan, Michael I. and Weiss, Yair},
title = {On spectral clustering: analysis and an algorithm},
year = {2001},
publisher = {MIT Press},
address = {Cambridge, MA, USA},
booktitle = {Proceedings of the 15th International Conference on Neural Information Processing Systems: Natural and Synthetic},
pages = {849–856},
numpages = {8},
location = {Vancouver, British Columbia, Canada},
series = {NIPS'01}
}

@article{bettini2022vmas,
  title = {VMAS: A Vectorized Multi-Agent Simulator for Collective Robot Learning},
  author = {Bettini, Matteo and Kortvelesy, Ryan and Blumenkamp, Jan and Prorok, Amanda},
  year = {2022},
  journal={The 16th International Symposium on Distributed Autonomous Robotic Systems},
  publisher={Springer}
}

@article{schulman2017proximal,
  title={Proximal policy optimization algorithms},
  author={Schulman, John and Wolski, Filip and Dhariwal, Prafulla and Radford, Alec and Klimov, Oleg},
  journal={arXiv preprint arXiv:1707.06347},
  year={2017}
}

@inproceedings{carroll2019utility,
 author = {Carroll, Micah and Shah, Rohin and Ho, Mark and Griffiths, Tom and Seshia, Sanjit and Abbeel, Pieter and Dragan, Anca},
 booktitle = {Advances in Neural Information Processing Systems},
 editor = {H. Wallach and H. Larochelle and A. Beygelzimer and F. d\textquotesingle Alch\'{e}-Buc and E. Fox and R. Garnett},
 pages = {},
 publisher = {Curran Associates, Inc.},
 title = {On the Utility of Learning about Humans for Human-AI Coordination},
 url = {https://proceedings.neurips.cc/paper_files/paper/2019/file/f5b1b89d98b7286673128a5fb112cb9a-Paper.pdf},
 volume = {32},
 year = {2019}
}

@inproceedings{
charakorn2023generating,
title={Generating Diverse Cooperative Agents by Learning Incompatible Policies},
author={Rujikorn Charakorn and Poramate Manoonpong and Nat Dilokthanakul},
booktitle={The Eleventh International Conference on Learning Representations },
year={2023},
url={https://openreview.net/forum?id=UkU05GOH7_6}
}

@article{cuiADVERSARIALDIVERSITYHANABI2023,
  title = {{{ADVERSARIAL DIVERSITY IN HANABI}}},
  author = {Cui, Brandon and Lupu, Andrei and Sokota, Samuel and Hu, Hengyuan and Wu, David J and Foerster, Jakob N},
  year = {2023},
  langid = {english}
}

@article{cuiSyKLRBRKlevelReasoning2021,
  title={K-level Reasoning for Zero-Shot Coordination in Hanabi},
  author={Brandon Cui and Hengyuan Hu and Luis Pineda and Jakob Nicolaus Foerster},
  journal={ArXiv},
  year={2022},
  volume={abs/2207.07166},
  url={https://api.semanticscholar.org/CorpusID:246996266}
}

@inproceedings{foersterLOLALearningOpponentLearning2018,
  title={Learning with Opponent-Learning Awareness},
  author={Jakob N. Foerster and Richard Y. Chen and Maruan Al-Shedivat and Shimon Whiteson and P. Abbeel and Igor Mordatch},
  booktitle={Adaptive Agents and Multi-Agent Systems},
  year={2017},
  url={https://api.semanticscholar.org/CorpusID:8708073}
}

@article{tampuu2017multiagent,
  title={Multiagent cooperation and competition with deep reinforcement learning},
  author={Tampuu, Ardi and Matiisen, Tambet and Kodelja, Dorian and Kuzovkin, Ilya and Korjus, Kristjan and Aru, Juhan and Aru, Jaan and Vicente, Raul},
  journal={PloS one},
  volume={12},
  number={4},
  pages={e0172395},
  year={2017},
  publisher={Public Library of Science San Francisco, CA USA}
}

@InProceedings{pmlr-v139-hu21c,
  title = 	 {Off-Belief Learning},
  author =       {Hu, Hengyuan and Lerer, Adam and Cui, Brandon and Pineda, Luis and Brown, Noam and Foerster, Jakob},
  booktitle = 	 {Proceedings of the 38th International Conference on Machine Learning},
  pages = 	 {4369--4379},
  year = 	 {2021},
  editor = 	 {Meila, Marina and Zhang, Tong},
  volume = 	 {139},
  series = 	 {Proceedings of Machine Learning Research},
  month = 	 {18--24 Jul},
  publisher =    {PMLR},
  url = 	 {https://proceedings.mlr.press/v139/hu21c.html}
}

@InProceedings{huOtherPlay2020,
  title = 	 {“{O}ther-Play” for Zero-Shot Coordination},
  author =       {Hu, Hengyuan and Lerer, Adam and Peysakhovich, Alex and Foerster, Jakob},
  booktitle = 	 {Proceedings of the 37th International Conference on Machine Learning},
  pages = 	 {4399--4410},
  year = 	 {2020},
  editor = 	 {III, Hal Daumé and Singh, Aarti},
  volume = 	 {119},
  series = 	 {Proceedings of Machine Learning Research},
  month = 	 {13--18 Jul},
  publisher =    {PMLR},
  url = 	 {https://proceedings.mlr.press/v119/hu20a.html}
}

@InProceedings{pmlr-v97-jaques19a,
  title = 	 {Social Influence as Intrinsic Motivation for Multi-Agent Deep Reinforcement Learning},
  author =       {Jaques, Natasha and Lazaridou, Angeliki and Hughes, Edward and Gulcehre, Caglar and Ortega, Pedro and Strouse, Dj and Leibo, Joel Z. and De Freitas, Nando},
  booktitle = 	 {Proceedings of the 36th International Conference on Machine Learning},
  pages = 	 {3040--3049},
  year = 	 {2019},
  editor = 	 {Chaudhuri, Kamalika and Salakhutdinov, Ruslan},
  volume = 	 {97},
  series = 	 {Proceedings of Machine Learning Research},
  month = 	 {09--15 Jun},
  publisher =    {PMLR},
  url = 	 {https://proceedings.mlr.press/v97/jaques19a.html}
}

@misc{luModelFreeOpponentShaping2022,
  title = {Model-{{Free Opponent Shaping}}},
  author = {Lu, Chris and Willi, Timon and de Witt, Christian Schroeder and Foerster, Jakob},
  year = {2022},
  month = nov,
  number = {arXiv:2205.01447},
  eprint = {2205.01447},
  primaryclass = {cs},
  publisher = {arXiv},
  doi = {10.48550/arXiv.2205.01447},
  urldate = {2025-01-20},
  archiveprefix = {arXiv}
}

@article{article,
author = {Camerer, Colin and Ho, Teck and Chong, Juin-Kuan},
year = {2004},
month = {01},
pages = {},
title = {A Cognitive Hierarchy Theory of One-shot Games}
}

@inproceedings{lupuTrajectoryDiversityZeroShot2021,
  title = {Trajectory {{Diversity}} for {{Zero-Shot Coordination}}},
  booktitle = {Proceedings of the 38th {{International Conference}} on {{Machine Learning}}},
  author = {Lupu, Andrei and Cui, Brandon and Hu, Hengyuan and Foerster, Jakob},
  year = {2021},
  month = jul,
  pages = {7204--7213},
  publisher = {PMLR},
  issn = {2640-3498},
  urldate = {2025-03-06},
  langid = {english}
}

@misc{mirsky2022surveyadhocteamwork,
      title={A Survey of Ad Hoc Teamwork Research}, 
      author={Reuth Mirsky and Ignacio Carlucho and Arrasy Rahman and Elliot Fosong and William Macke and Mohan Sridharan and Peter Stone and Stefano V. Albrecht},
      year={2022},
      eprint={2202.10450},
      archivePrefix={arXiv},
      primaryClass={cs.MA},
      url={https://arxiv.org/abs/2202.10450}, 
}

@book{oliehoekConciseIntroductionDecentralized2016,
  title = {A {{Concise Introduction}} to {{Decentralized POMDPs}}},
  author = {Oliehoek, Frans A. and Amato, Christopher},
  year = {2016},
  series = {{{SpringerBriefs}} in {{Intelligent Systems}}},
  publisher = {Springer International Publishing},
  address = {Cham},
  doi = {10.1007/978-3-319-28929-8},
  urldate = {2024-11-01},
  copyright = {http://www.springer.com/tdm},
  isbn = {978-3-319-28927-4 978-3-319-28929-8},
  langid = {english}
}

@misc{sarkarComediDiverseConventions2023,
  title = {Comedi -- {{Diverse Conventions}} for {{Human-AI Collaboration}}},
  author = {Sarkar, Bidipta and Shih, Andy and Sadigh, Dorsa},
  year = {2023},
  month = oct,
  number = {arXiv:2310.15414},
  eprint = {2310.15414},
  primaryclass = {cs},
  publisher = {arXiv},
  doi = {10.48550/arXiv.2310.15414},
  urldate = {2025-05-21},
  archiveprefix = {arXiv}
}

@inproceedings{10.1145/3306618.3314268,
author = {Lerer, Adam and Peysakhovich, Alexander},
title = {Learning Existing Social Conventions via Observationally Augmented Self-Play},
year = {2019},
isbn = {9781450363242},
publisher = {Association for Computing Machinery},
address = {New York, NY, USA},
url = {https://doi.org/10.1145/3306618.3314268},
doi = {10.1145/3306618.3314268},
booktitle = {Proceedings of the 2019 AAAI/ACM Conference on AI, Ethics, and Society},
pages = {107–114},
numpages = {8},
location = {Honolulu, HI, USA},
series = {AIES '19}
}

@inproceedings{DBLP:conf/iclr/ShihSKES21,
  author       = {Andy Shih and
                  Arjun Sawhney and
                  Jovana Kondic and
                  Stefano Ermon and
                  Dorsa Sadigh},
  title        = {On the Critical Role of Conventions in Adaptive Human-AI Collaboration},
  booktitle    = {9th International Conference on Learning Representations, {ICLR} 2021,
                  Virtual Event, Austria, May 3-7, 2021},
  publisher    = {OpenReview.net},
  year         = {2021},
  url          = {https://openreview.net/forum?id=8Ln-Bq0mZcy},
  bibsource    = {dblp computer science bibliography, https://dblp.org}
}

@article{Camerer2003,
  title = {A Cognitive Hierarchy Theory of One-shot Games and Experimental Analysis},
  ISSN = {1556-5068},
  url = {http://dx.doi.org/10.2139/ssrn.411061},
  DOI = {10.2139/ssrn.411061},
  journal = {SSRN Electronic Journal},
  publisher = {Elsevier BV},
  author = {Camerer,  Colin F. and Ho,  Teck and Chong,  Juin-Kuan},
  year = {2003}
}

@inproceedings{strouseFCPCollaboratingHumans2021,
 author = {Strouse, DJ and McKee, Kevin and Botvinick, Matt and Hughes, Edward and Everett, Richard},
 booktitle = {Advances in Neural Information Processing Systems},
 editor = {M. Ranzato and A. Beygelzimer and Y. Dauphin and P.S. Liang and J. Wortman Vaughan},
 pages = {14502--14515},
 publisher = {Curran Associates, Inc.},
 title = {Collaborating with Humans without Human Data},
 url = {https://proceedings.neurips.cc/paper_files/paper/2021/file/797134c3e42371bb4979a462eb2f042a-Paper.pdf},
 volume = {34},
 year = {2021}
}

@ARTICLE{zhaoMEPMaximumEntropy2022,
  title     = "Maximum Entropy {Population-Based} Training for zero-shot
               {human-AI} coordination",
  author    = "Zhao, Rui and Song, Jinming and Yuan, Yufeng and Hu, Haifeng and
               Gao, Yang and Wu, Yi and Sun, Zhongqian and Yang, Wei",
  journal   = "Proc. Conf. AAAI Artif. Intell.",
  publisher = "Association for the Advancement of Artificial Intelligence
               (AAAI)",
  volume    =  37,
  number    =  5,
  pages     = "6145--6153",
  month     =  jun,
  year      =  2023
}

@article{Stone_Kaminka_Kraus_Rosenschein_2010, title={Ad Hoc Autonomous Agent Teams: Collaboration without Pre-Coordination}, volume={24}, url={https://ojs.aaai.org/index.php/AAAI/article/view/7529}, DOI={10.1609/aaai.v24i1.7529}, abstractNote={ &lt;p&gt; As autonomous agents proliferate in the real world, both in software and robotic settings, they will increasingly need to band together for cooperative activities with previously unfamiliar teammates. In such ad hoc team settings, team strategies cannot be developed a priori. Rather, an agent must be prepared to cooperate with many types of teammates: it must collaborate without pre-coordination. This paper challenges the AI community to develop theory and to implement prototypes of ad hoc team agents. It defines the concept of ad hoc team agents, specifies an evaluation paradigm, and provides examples of possible theoretical and empirical approaches to challenge. The goal is to encourage progress towards this ambitious, newly realistic, and increasingly important research goal. &lt;/p&gt; }, number={1}, journal={Proceedings of the AAAI Conference on Artificial Intelligence}, author={Stone, Peter and Kaminka, Gal and Kraus, Sarit and Rosenschein, Jeffrey}, year={2010}, month={Jul.}, pages={1504-1509} }

@article{scikit-learn,
  title={Scikit-learn: Machine Learning in {P}ython},
  author={Pedregosa, F. and Varoquaux, G. and Gramfort, A. and Michel, V.
          and Thirion, B. and Grisel, O. and Blondel, M. and Prettenhofer, P.
          and Weiss, R. and Dubourg, V. and Vanderplas, J. and Passos, A. and
          Cournapeau, D. and Brucher, M. and Perrot, M. and Duchesnay, E.},
  journal={Journal of Machine Learning Research},
  volume={12},
  pages={2825--2830},
  year={2011}
}

@inproceedings{10.65109/UMCV8852,
author = {Ni, Andrew and Stepputtis, Simon and Nikolaidis, Stefanos and Lewis, Michael and Sycara, Katia P. and Kim, Woojun},
title = {Theory of Mind Guided Strategy Adaptation for Zero-Shot Coordination},
year = {2026},
isbn = {9798400723179},
publisher = {International Foundation for Autonomous Agents and Multiagent Systems},
address = {Richland, SC},
url = {https://doi.org/10.65109/UMCV8852},
doi = {10.65109/UMCV8852},
booktitle = {Proceedings of the 25th International Conference on Autonomous Agents and Multiagent Systems},
pages = {1810–1819},
numpages = {10},
location = {Paphos, Cyprus},
series = {AAMAS '26}
}
